\documentclass[10pt]{IEEEtran}
\usepackage{cite}
\usepackage{amsmath,amssymb,amsfonts}
\usepackage{graphicx}
\usepackage{textcomp}
\usepackage{xcolor}
\def\BibTeX{{\rm B\kern-.05em{\sc i\kern-.025em b}\kern-.08em
    T\kern-.1667em\lower.7ex\hbox{E}\kern-.125emX}}

\usepackage{url}
\usepackage{subcaption}
\usepackage{booktabs}
\usepackage{multirow}
\usepackage{tabularx}
\usepackage{enumitem}

\usepackage{glossaries}
\newacronym{MLR}{MLR}{multi-vocal literature review}
\newacronym{ICT}{ICT}{information and communication technology}
\newacronym{LLM}{LLM}{large language model}
\newacronym{RL}{RL}{reinforcement learning}
\newacronym{EDP}{EDP}{energy delay product}
\newacronym{SLA}{SLA}{service-level agreement}
\newacronym{DCCCE}{DCCCE}{data center and cloud computing environment}
\newacronym{VM}{VM}{virtual machine}
\newacronym{CO2e}{CO2e}{carbon dioxide equivalent}
\newacronym{RQ}{RQ}{research question}
\newacronym{EE}{EE}{energy efficiency}
\newacronym{PC}{PC}{power consumption}
\newacronym{CoT}{CoT}{chain-of-thought}
\newacronym{ICL}{ICL}{in-context learning}
\newacronym{PM}{PM}{physical machine}
\newacronym{DVFS}{DVFS}{dynamic voltage and frequency scaling}
\newacronym{FLOP}{FLOP}{floating-point operation}
\newacronym{GSF}{GSF}{Green Software Foundation}
\newacronym{AI}{AI}{artificial intelligence}
\newacronym{HPC}{HPC}{high-performance computing}
\newacronym{PDU}{PDU}{power distribution unit}
\newacronym{IPMI}{IPMI}{intelligent platform management interface}
\newacronym{RAPL}{RAPL}{Running Average Power Limit}
\newacronym{IOPS}{IOPS}{input/output operations}
\newacronym{SWaP}{SWaP}{space, watts, and performance}
\newacronym{DH-UR}{DH-UR}{deployed hardware utilization ratio}
\newacronym{DH-UE}{DH-UE}{deployed hardware utilization efficiency}
\newacronym{PUE}{PUE}{power usage effectiveness}
\newacronym{DCIE}{DCIE}{data center infrastructure efficiency}
\newacronym{CPE}{CPE}{compute power efficiency}
\newacronym{DC-EEP}{DC-EEP}{data center energy efficiency and productivity}
\newacronym{SI-EER}{SI-EER}{site infrastructure energy efficiency ratio}
\newacronym{SI-POM}{SI-POM}{site infrastructure power overhead multiplier}
\newacronym{H-POM}{H-POM}{hardware power overhead multiplier}
\newacronym{DCeP}{DCeP}{data center energy productivity}
\newacronym{DCPE}{DCPE}{data center performance efficiency}
\newacronym{COP}{COP}{coefficient of performance}
\newacronym{CUE}{CUE}{carbon usage effectiveness}
\newacronym{GPUE}{GPUE}{green power usage effectiveness}
\newacronym{SCI}{SCI}{software carbon intensity}

\usepackage[breaklinks,hidelinks]{hyperref}

\begin{document}

\title{Measuring and Optimizing Energy Efficiency in ICT: A Multi-vocal Literature Review on Data Centers, Cloud Computing, and LLM-Driven Code Optimization}

\author{
\IEEEauthorblockN{
Tobias Böhm\textsuperscript{1},
Benjamin Weigell\textsuperscript{2},
Ezgi Sarikayak\textsuperscript{3,4},
Philipp M. Zähl\textsuperscript{5},
Tina Vartziotis\textsuperscript{6,7},
Simon Hornung\textsuperscript{2},
Julien Siebert\textsuperscript{8},
Anika Hennig\textsuperscript{5},
Bernhard Bauer\textsuperscript{2},
Andreas Biesdorf\textsuperscript{1,4}
}
\\[0.7em]
\IEEEauthorblockA{
\textsuperscript{1}Trier University of Applied Sciences, Trier, Germany
\textsuperscript{2}University of Augsburg, Augsburg, Germany
\textsuperscript{3}Technical University of Munich, Munich, Germany
\textsuperscript{4}Siemens AG, Munich, Germany
\textsuperscript{5}FH Aachen University of Applied Sciences, Aachen, Germany
\textsuperscript{6}TWT Science \& Innovation, Stuttgart, Germany
\textsuperscript{7}National Technical University of Athens, Athens, Greece
\textsuperscript{8}Fraunhofer IESE, Kaiserslautern, Germany
}
\thanks{Corresponding author: Tobias Böhm (t.boehm@wir.hochschule-trier.de)}
}

\maketitle

\begin{abstract}
Rapid growth in Information and Communication Technology (ICT) increases electricity use and environmental impact. 
Addressing this challenge requires a systematic, software-centric view of how energy and related sustainability attributes are measured, reported, and improved. 
We present a comprehensive Multi-vocal Literature Review (MLR) integrating peer-reviewed and gray literature to map current practice in measuring and optimizing energy use in ICT software, with emphasis on \glspl{DCCCE}.
First, we synthesize how sustainability is \emph{measured} across ICT by proposing a taxonomy organized along \emph{metric domains} (power/energy, energy efficiency, environmental impact) and \emph{component granularity} (from applications, to containers/VMs, and system-of-systems levels). 
We summarize measurement approaches and the tooling landscape that implements them. 
Second, we review \emph{optimization} strategies at the software and system levels, covering dynamic power and resource management, VM/container consolidation, energy-aware software and system design, and cross-layer, system-wide methods relevant to \glspl{DCCCE}.
Third, we examine how large language models (LLMs) are applied to \emph{code generation and transformation} for resource-efficient software, noting that most studies target code generation (predominantly in Python) and evaluate runtime and memory, with fewer works directly reporting energy.
Hardware- and network-level telemetry are treated as data sources for software-level analysis. 
We identify gaps in vendor-neutral telemetry, consistent measurement and reporting practices, network-level metrics, and unified, cross-layer optimization frameworks, and outline directions for improving reproducibility and comparability. 
We screened 635 records (2010-2025) and included 181 sources (122 peer-reviewed, 59 gray).
\end{abstract}
\glsreset{DCCCE}


\section{Introduction}
\label{introduction}
The global volume of \gls{ICT} has grown rapidly over the past decade, driven particularly by data centers, cloud computing, and industrial digitalization. This growth comes at a rising environmental cost, since it increases electricity use and, in turn, greenhouse gas emissions. As energy consumption in \gls{ICT} is projected to continue to rise, the need for sustainable \gls{ICT} practices with more focus on measurement and efficiency is now critical~\cite{koot_usage_2021}.

The field of Green Software Engineering addresses these challenges by making energy a first-class concern, measuring and improving it at the software level. Despite numerous efforts, the current landscape of tools, methods, and approaches remains fragmented. Prior work uses heterogeneous measurement targets and units (energy, power, carbon), reports different baselines and granularities, and relies on tools with limited external validation, hampering comparability and reproducibility. What is missing is a coherent, software-centric synthesis, namely, a taxonomy and evidence map that link metrics, approaches, and developer-facing tools, especially for \glspl{DCCCE}.

This review provides that synthesis. We (1) synthesize and classify state-of-the-art approaches, methods, tools, and metrics for optimizing energy efficiency and power consumption with a thematic focus on \glspl{DCCCE}; (2) analyze how these works operationalize key measurement constructs (targets and units); and (3) examine the emerging role of \glspl{LLM} in automating code generation and transformation toward resource-efficient software. Physical data center design and layout are explicitly out of scope. Instead, we focus on software behavior, software architecture, and developer-facing tooling. We provide a reproducible synthesis that informs classification, quantification, and thematic groupings across approaches, methods, tools, and metrics. Section~\ref{methodology} details the search, screening, and synthesis protocols.

\subsection{Structure of the Article}
The remainder of this article is organized as follows: Section~\ref{background} reviews the landscape and related work. Section~\ref{methodology} presents the research questions and the multi-vocal literature review design. Section~\ref{results} reports the findings. Section~\ref{discussion} discusses them in light of the research questions and outlines limitations and threats to validity. Finally, Section~\ref{conclusion} summarizes the contributions and points to future work.

\section{Background and Related Work}
\label{background}
\subsection{Current Landscape of Related Research}
Green IT, also referred to as sustainable or green computing, aims to reduce the environmental footprint of \gls{ICT} systems across their life cycle, primarily by improving energy efficiency~\cite{kipp_applying_2012,priya_survey_2013}. Over the past 15 years, scientific interest has grown substantially, with contributions spanning data centers and cloud/edge platforms~\cite{kliazovich_greencloud_2012,khattar_toward_2019}, IoT and embedded systems~\cite{albreem_green_2017}, AI/ML workloads~\cite{barbierato_toward_2024}, and software engineering methods and tools~\cite{meinhardt_software_2023,balanza-martinez_tactics_2024}. In this review, we adopt a \textit{software-centric} perspective, with emphasis on \glspl{DCCCE}.

The literature can be broadly grouped into several established subfields: green data center research focuses on energy-efficient design and operation of data centers, sustainable cooling, integration of renewable energies, and hardware circularity/waste reduction. Green cloud computing emphasizes reducing the environmental impact of cloud services through efficient virtualization, optimized resource allocation, and increased use of sustainable energy sources~\cite{priya_survey_2013}. Green IoT seeks energy-efficient IoT architectures, in particular via low-power hardware, local data processing, and intelligent energy management~\cite{albreem_green_2017}. Green AI investigates how to minimize the ecological footprint of AI applications, including \glspl{LLM}, by monitoring and reducing energy use, optimizing hyperparameters, and developing efficient benchmarking methods~\cite{barbierato_toward_2024}. Finally, research on energy-efficient hardware and software addresses low-power processors, more durable storage technologies (e.g., SSDs), and software-level algorithmic efficiency~\cite{paul_comprehensive_2023}. 

Within software-based Green IT research, approaches range from designing resource-efficient software that improves infrastructure utilization via virtualization and cloud platforms to embedding IT practices within broader sustainability policies to extend device lifetimes. Current research clusters around two major themes:  (1) 
classification and measurement of power and energy consumption (including targets, units, baselines, and granularity), and (2) 
methods for optimizing energy efficiency and power consumption at the software and system levels. Of particular relevance to this review, as well as the GreenCode research project that informs it, are methods focused on \glspl{DCCCE} and the use of \glspl{LLM} for code generation and transformation toward improved resource efficiency.

There are numerous systematic literature reviews covering these subfields. For example, Ergasheva et al.~\cite{ergasheva_metrics_2020} and Bozzelli et al.~\cite{bozzelli_systematic_nodate} focus on both general and software-specific green metrics for assessing energy consumption in software systems. Furthermore, Balanza-Martinez et al.~\cite{balanza-martinez_tactics_2024} and Wysocki et al.~\cite{wysocki_methods_2025} survey and compare optimization approaches aimed at improving energy efficiency in software systems. Connolly Bree and Ó Cinnéide~\cite{connolly_bree_how_2025} extend this line of work by examining how software design decisions shape system-level performance. Studies on the resource footprint of \glspl{LLM} in software systems also exist, but to our knowledge none evaluate  their direct impact on software energy efficiency (e.g., end-to-end energy savings from LLM-based code generation or transformation)~\cite{shi_efficient_2025,hou_large_2024}.

Reviewing this body of work reveals clear gaps: Existing systematic literature reviews analyze individual subfields in detail, but a holistic, cross-cutting perspective is largely absent. Furthermore, prior reviews generally exclude non-peer-reviewed and industrial contributions outside of traditional publishing channels (gray literature) and concentrate only on academic, peer-reviewed publications. Accordingly, this review contributes a systematic, software-centric, and cross-cutting synthesis that integrates both academic and gray literature. Consistent with our scope, the focus is on software behavior, architecture, and developer-facing tooling in \glspl{DCCCE}.\ Hardware- and network-level telemetry are incorporated as data sources rather than design targets. We further examine the emerging role of \glspl{LLM} for resource-efficiency in code generation and transformation, as prior work on Green AI rarely evaluates end-to-end effects on software energy efficiency. By systematizing these threads, we address a central research gap and provide a timely, comprehensive reference for measurement and optimization practice in Green IT.

\section{Methodology}
\label{methodology}
This section describes our research methodology and outlines the systematic approach used to gather literature for this study. We conducted a Multi-vocal Literature Review (MLR) following the systematic process defined by Garousi et al.~\cite{garousi_guidelines_2019}, with the goal producing a comprehensive synthesis of the topic and integrating relevant insights across sources. This approach does not limit the study to scientifically published journal articles, books, or conference proceedings, but also incorporates gray literature, including technical reports, white papers, preprints, and lecture materials. By systematically examining these sources, the \gls{MLR} aims to identify relevant connections, knowledge gaps, inconsistencies, and unexplored areas that might be overlooked in narrower literature reviews. This approach ensures a balanced and objective analysis, capturing both detailed viewpoints and broader, systemic perspectives of the state-of-the-art in Green IT. The remainder of this section details the applied methodology, including search strategy, screening and inclusion criteria, quality assessment for gray literature, data extraction, and synthesis procedures.

\subsection{Research Goals and Research Questions}
Within the broader topic of Green IT, this study aims to provide a structured and holistic overview of the state-of-the-art and practice in measuring and optimizing power and energy use in \gls{ICT} systems. Of particular interest are software- and system-level approaches, methods, metrics, and tools, and how they can be classified and thematically grouped. We also examine the emerging role of \glspl{LLM} in resource-efficient code generation and transformation. Accordingly, we formulate three \glspl{RQ} (see Table \ref{tab:RQs}).

\begin{table}[htbp]
\caption{Research Questions (RQs)}
\centering
\begin{tabular}{cp{6cm}}
\toprule
\textbf{\#} & \textbf{Research Question} \\
\toprule
\hypertarget{RQ1}{RQ1} & How is sustainability (energy, power, carbon) quantified in \gls{ICT} systems, and how can the resulting metrics be classified (e.g., by target, level, and method)? \\
\midrule
\hypertarget{RQ2}{RQ2} & Which software- and system-level approaches, methods, tools, and techniques improve energy efficiency and reduce power/energy use in \gls{ICT} systems, specifically in data centers and cloud computing environments (DCCCEs)? \\
\midrule
\hypertarget{RQ3}{RQ3} & How are \glspl{LLM} used to automate code generation and transformation toward resource-efficient software, and how is their effectiveness evaluated (e.g., runtime, memory, energy)?\\
\bottomrule
\end{tabular}
\label{tab:RQs}
\end{table}

\subsection{Search Process and Source Selection}
\begin{figure*}[htbp]
\centerline{\includegraphics[width=0.82\linewidth]{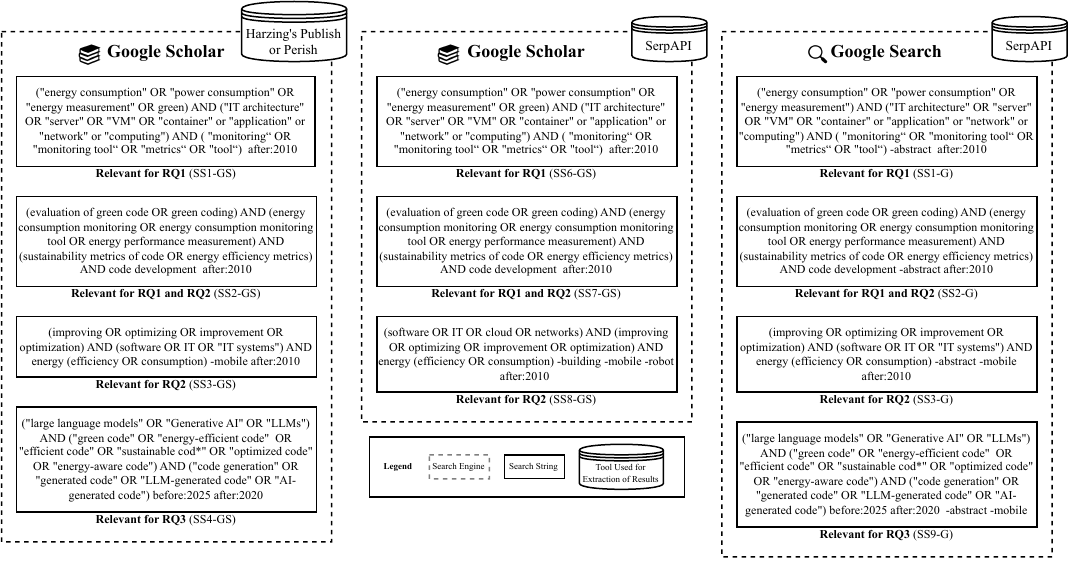}}
\caption{Overview of Search Terms}
\label{fig:search_strings}
\end{figure*}

The search strategy used for this \gls{MLR} followed a systematic construction of search terms that combine relevant concepts and synonyms. The search terms were chosen to cover the central aspects of our \glspl{RQ} and can be found in Fig.~\ref{fig:search_strings}. In addition, we applied a time window (2010–2025) to capture contemporary work on cloud and software energy measurement/optimization. LLM-specific queries were further constrained to 2020–2025. An initially considered research question on legacy system migration was later dropped for follow-up to preserve thematic coherence in this study.

\begin{figure}[htbp]
\centerline{\includegraphics[width=0.66\linewidth]{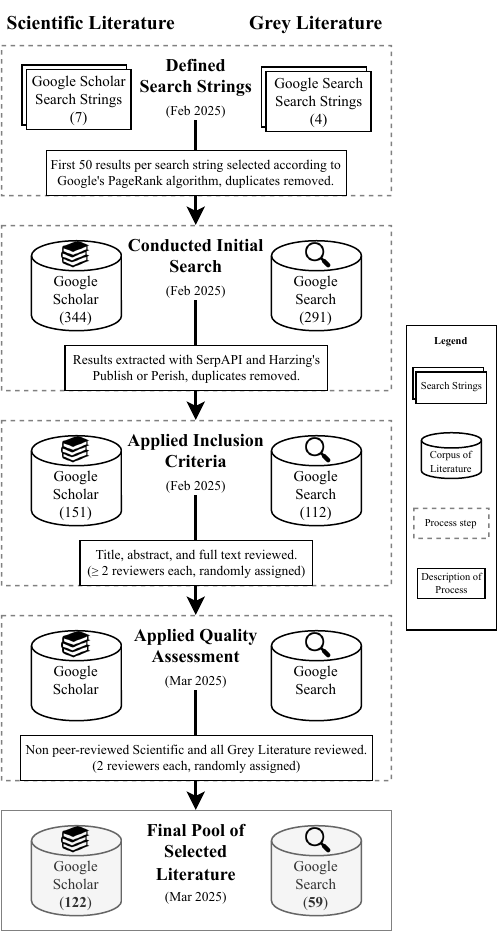}}
\caption{Steps of the Search Process}
\label{fig:search_process}
\end{figure}

An overview of the search process is outlined in Fig. \ref{fig:search_process}. The initial search identified 635 sources: 344 from Google Scholar and 291 from Google Search retrieved via SerpAPI and Harzing's Publish or Perish, then deduplicated. Next, we applied the inclusion criteria defined in 
Table~\ref{tab:inclusion_criteria}: title, abstracts, and (where accessible) full texts were reviewed by at least two reviewers, working independently with randomized assignment to reduce bias. 
Sources considered unsuitable were excluded. Finally, sources that met the inclusion criteria and were either non-peer-reviewed scientific or gray literature were additionally assessed using the quality criteria in Table~\ref{tab:qual_assessment_crit}. In case of disagreement between two reviewers, a third reviewer served to make the final decision. The final pool of selected literature comprises 181 sources, 122 scientific and 59 gray literature sources (see Fig.~\ref{fig:search_process}).

Restricting Google Search and Google Scholar results to the first 50 hits per query (Table~\ref{tab:inclusion_criteria}, IC4-IC5) reflects a pragmatic, fixed stopping rule for an otherwise unbounded gray-literature search space, consistent with practical MLR search guidance~\cite{garousi_guidelines_2019}; we discuss the resulting bias and coverage trade-offs in Section~\ref{limitations-threats-validity}.

\begin{table}[htbp]
\caption{List of Inclusion Criteria (IC)}
\centering
\begin{tabular}{cp{6cm}}
\toprule
\textbf{\#} & \textbf{Inclusion Criterion} \\
\toprule
IC1 & In case of ambiguities or disagreements, a third reviewer resolves disputes (2/3 agreement). \\
\midrule
IC2 & Must be in English. \\
\midrule
IC3 & Must be published in 2010 to 2025; LLM-specific sources must be published in 2020 to 2025. \\
\midrule
IC4 & Sourced from Google Search (first 50 results). \\
\midrule
IC5 & Sourced from Google Scholar (first 50 results). \\
\midrule
IC6 & Relevant to the research questions (RQs). \\
\midrule
IC7 & Full text accessible. \\
\midrule
IC8 & Acceptable formats: books, magazines, theses, government reports, white papers, news articles, videos, Q\&A sites (e.g., StackOverflow), wikis, blog posts, presentations, emails, tweets. \\
\midrule
IC9 & Non peer-reviewed scientific and gray literature must pass a quality assessment (Table~\ref{tab:qual_assessment_crit}). Peer-reviewed items do not undergo this additional QA screen. \\
\bottomrule
\end{tabular}
\label{tab:inclusion_criteria}
\end{table}

\begin{table}[htbp]
\caption{List of Quality Assessment Criteria (QC)}
\centering
\begin{tabular}{ccp{4cm}}
\toprule
\textbf{\#} & \textbf{Category} & \textbf{Quality Criterion} \\
\toprule
QC1 & Date & Does it have a clearly stated date? \\
\midrule
QC2 & Authorship & Is an individual author or organization associated with it? \\
\midrule
QC3 & Authorship & Does the author or organization have other published work in the field? \\
\midrule
QC4 & Methodology & Does it describe a clearly stated methodology, approach, metrics, or tool? \\
\midrule
QC5 & Relevance & Does it cover a specific question relevant to the RQs in this review? \\
\bottomrule
\end{tabular}
\label{tab:qual_assessment_crit}
\end{table}

\subsection{Data Extraction}
The final sources were systematically analyzed and structured for data extraction. We used a pre-defined codebook and a standardized extraction form to consistently capture relevant information across all included items. The form recorded bibliographic data, source type (peer reviewed or gray), context (application domain and \gls{DCCCE} relevance), methodological details, measurement targets and units, baseline definitions, reporting granularity, metrics, tools, datasets or benchmarks, evaluation criteria and outcomes, threats to validity as reported by the authors, and the mapping to RQ1 to RQ3.  Two reviewers independently piloted the form on a random subset of ten items each, reconciled differences to refine the codebook, and then proceeded with single extraction plus targeted verification on ambiguous fields. Disagreements were resolved by a third reviewer. All extractions were maintained in a shared spreadsheet with version history to ensure traceability. This systematic approach ensured an objective and transparent preparation of the data, supporting robust analysis and synthesis of the existing literature.

\subsection{Data Synthesis \& Analysis}
A narrative approach to data synthesis and analysis was chosen to qualitatively structure and interpret the extracted information. We organized studies along the axes defined by the RQs (measurement/classification, optimization in DCCCEs, and LLM-enabled code generation/transformation) and applied an iterative thematic coding process to group comparable evidence. This approach allows for a detailed and contextualized presentation of the identified research and practice sources. The narrative synthesis highlighted key findings on existing approaches, methods, metrics and tools for energy measurement and optimization in \gls{ICT} systems, examines the role of \glspl{LLM} in resource-efficient code generation and transformation, and identifies relevant connections between the different studies within these main topics.

\section{Results}
\label{results}
The following sections report the \gls{MLR} results organized by the guiding \glspl{RQ}. After  data extraction, we synthesized the material into thematically coherent groups to improve comparability across diverse sources. Findings are presented in three categories, aligned with RQ1-RQ3: \ref{ee-measurement} Classification and Measurement of Sustainability Metrics in \gls{ICT} Systems (19 peer-reviewed and 8 gray); \ref{ee-optimization} Software- and System-Level Optimization of Power Consumption and Energy Efficiency (22 peer-reviewed; 3 gray); and \ref{ee-code-optimization} Code Generation and Transformation for Resource-Efficient Software (19 peer-reviewed; 2 gray). Of the 181 included sources, these 73 are discussed individually in the narrative synthesis below; the remainder informed the taxonomy, classification, and background framing without individual citation. Gray literature is proportionally less represented among the 73 individually discussed sources (13 of 73) than among the full included pool (59 of 181): consistent with our quality-assessment screening (Table~\ref{tab:qual_assessment_crit}), gray sources more often contributed contextual and tooling-landscape information than material central to the narrative synthesis (see Section~\ref{limitations-threats-validity} for further discussion).

\subsection{Classification and Measurement of Sustainability Metrics in ICT Systems}
\label{ee-measurement}

We address RQ1 by analyzing how sustainability in \gls{ICT} systems is measured across academic and practitioner sources. For scope, we consider three component types: (i) hardware (servers, CPUs, and networking equipment); (ii) software at two granularities: whole elements (VMs, containers, operating systems, applications), and partial elements (functions, routines, code snippets); and (iii) inter-component communication. In Section~\ref{ee-measurement-classification} we derive, from 27 sources, a taxonomy with three metric domains: power/energy, energy efficiency, and environmental impact. In Section~\ref{ee-measurement-approaches-tooling} we summarize common measurement approaches (external meters, on-device sensing, model-based estimation) and the supporting tools. 

\subsubsection{Classification of Sustainability Metrics}
\label{ee-measurement-classification}
We classify sustainability metrics in \gls{ICT} systems into three interrelated domains: \ref{ee-measurement-classification-energy-consumption} power and energy consumption, \ref{ee-measurement-classification-energy-efficiency} energy efficiency, and \ref{ee-measurement-classification-environmental-impact} environmental impact. These domains offer complementary views -- raw resource use, use per unit of output, and broader consequences -- and together provide a coherent basis for measurement and comparison.

Power and energy metrics underpin sustainability assessment at both the hardware and software levels. Power (W) is the instantaneous rate of consumption, while energy (J, kWh) accumulates over time (the integral of power). These baselines enable further analysis and comparison. Efficiency metrics normalize consumption by useful work (e.g., throughput, operations, or computational work), such as FLOPS/W or transactions/kWh, allowing like-for-like comparisons across systems and configuration.  Environmental-impact metrics map energy use to broader consequences (e.g., \gls{CO2e}) using grid carbon intensity and, where available, embodied emissions, thereby situating ICT energy use within a wider sustainability context.

In the following, we detail each domain and highlight representative  metrics, measurement approaches, and tools.

\paragraph{Power and Energy Consumption Metrics}
\label{ee-measurement-classification-energy-consumption}
Power and energy consumption metrics form the empirical basis for sustainability assessment in \gls{ICT}. They quantify the raw amount of electricity used at an instant (power, W) or over an interface (energy, J or kWh) and thus underpin subsequent analysis and comparison~\cite{kipp_applying_2012,priya_survey_2013}. Accurate measurement is essential for reliable insights. We distinguish two primary loci of application: (i) the \emph{hardware level} (from racks/servers to components such as CPUs, memory, storage, and accelerators) and (ii) the \emph{software level} (from whole applications, \glspl{VM}, and containers down to functions and routines), with some work attributing energy to communication between software elements.

At the \emph{hardware level}, measurement targets the physical components that draw power. Work typically starts at a coarse granularity -- whole machines, such as servers, racks, routers, and switches -- and then drills down to individual components~\cite{bergen_towards_2014,kipp_applying_2012,kipp_layered_2012,uchechukwu_energy_2014,jiang_edge_2019,chen_profiling_2011}. 
The CPU is the most frequently profiled unit~\cite{noureddine_review_2013,uchechukwu_energy_2014,drraghavendra_holistic_2024,priya_survey_2013,jiang_edge_2019,chen_profiling_2011}, often alongside the motherboard~\cite{noureddine_review_2013,drraghavendra_holistic_2024} and main memory~\cite{noureddine_review_2013,drraghavendra_holistic_2024,priya_survey_2013,chen_profiling_2011}. Storage (HDDs/SSDs) is also a common focus, reflecting its nontrivial share of overall consumption~\cite{uchechukwu_energy_2014,drraghavendra_holistic_2024,priya_survey_2013,chen_profiling_2011,jiang_edge_2019}. 

With the growing adoption of machine learning and data-intensive workloads, hardware accelerators, especially GPUs, are increasingly profiled for power and energy~\cite{guldner_development_2024}.

Several studies also measure cooling subsystems (e.g., server fans, CRAC/air conditioning), given their significant share of data center energy use~\cite{uchechukwu_energy_2014,drraghavendra_holistic_2024,jiang_edge_2019}.

At the \emph{software level}, power and energy can be attributed to software entities at multiple granularities. Following our \gls{ICT} system model, we distinguish (i) \emph{full} software elements, such as complete programs and  virtualized environments (e.g., \glspl{VM}, containers), and (ii) \emph{partial} elements, such as individual services, routines, and functions. This split is reflected in the literature: energy is assessed for \glspl{VM}~\cite{guldner_development_2024,naqi_ai_nodate}, containers~\cite{guldner_development_2024,naqi_ai_nodate,noureddine_review_2013}, and whole applications/executables~\cite{noauthor_q-rzenamel_nodate,guldner_development_2024}. Studies also differentiate by application context (e.g., business process-driven systems vs. \gls{HPC} applications)~\cite{kipp_applying_2012}. Software-level attribution enables energy-aware decisions about deployment, resource allocation, migration, and scheduling. In practice, software energy is typically \emph{inferred} by mapping measured hardware energy to software components via attribution models; we review these models in Section~\ref{ee-measurement-approaches-tooling}.

A concise summary of how these metrics apply across system components and granularities is provided in Table~\ref{tab:power-energy-classification}.

\begin{table*}[ht]
\centering
\renewcommand{\arraystretch}{1.0}
\caption{Power and Energy Metrics by System Element and Granularity}
\label{tab:power-energy-classification}
\begin{tabular}{@{}llp{6.5cm}ll@{}}
\toprule
\textbf{Metric Type} & \textbf{Unit} & \textbf{Target Element} & \textbf{Granularity} & \textbf{Application Level} \\ \toprule
Power                & W (Watt)      & Entire machine (server, rack, etc.) & Coarse-grained & Hardware \\ \midrule
Energy               & J, kWh        & Entire machine (server, rack, etc.) & Coarse-grained & Hardware \\ \midrule
Power / Energy       & W / J / kWh   & CPU, Memory, Motherboard, Storage (HDD, SSD), GPU / Hardware Accelerators, Cooling systems (fans, AC) & Medium-grained & Hardware \\ \midrule
Energy               & J / kWh       & Virtual Machines (VMs), Containers, Full Applications / Executables & Full Software Element & Software \\ \midrule
Energy               & J / kWh       & Functions / Services / Routines & Partial Software Element & Software \\
\bottomrule
\end{tabular}
\end{table*}

\paragraph{Energy and Power Efficiency Metrics}
\label{ee-measurement-classification-energy-efficiency}
Efficiency metrics relate a system's useful work to the energy or power required to deliver it. They build on the power and energy measures introduced above and enable comparisons across configurations, implementations, and platforms, provided that workload and functional unit are specified consistently. Depending on the focus, energy efficiency can be expressed using either energy (e.g., J, kWh) or power (e.g., W) measurements.

The reviewed literature presents a diverse set of hardware-oriented efficiency metrics grouped into the following categories: (i) \emph{component-level} efficiency metrics, (ii) \emph{throughput-based} metrics, and (iii) \emph{system of systems} metrics (Table \ref{tab:energy-efficiency-metrics}).

\emph{Component-level} metrics quantify how efficiently individual components convert supplied energy into work, for example, by relating component energy to utilization or activity levels (e.g., CPU or memory power per utilization)~\cite{chen_profiling_2011, beik_green_2012}. 

\emph{Throughput-based} metrics express delivered work per unit of power, such as transactions per kilowatt-hour,  bandwidth per watt, or storage capacity per watt~\cite{kipp_applying_2012,kipp_layered_2012}.
These measures can be generalized to operations-centric indicators such as \gls{IOPS} per watt, capturing the number of discrete operations performed per unit of power~\cite{kipp_layered_2012}.

Finally, \emph{system-of-systems} metrics characterize the efficiency of entire facilities, such as data centers or cloud sites, by relating total facility power to IT power and, where possible, to delivered compute work. Common examples include \gls{PUE}, \gls{CPE}, and the \gls{DC-EEP} index~\cite{kipp_layered_2012}. These metrics capture infrastructure overheads (e.g., power distribution and cooling) and are therefore essential when evaluating end-to-end \glspl{DCCCE} efficiency.

At the software level, energy efficiency is typically assessed comparatively by running a fixed workload on a controlled platform and reporting the energy-performance trade-off between alternative implementations. A widely used composite indicator is the \gls{EDP}, defined as the product of task energy and execution time~\cite{green_software_foundation_gps-up_2021}. While \gls{EDP} rewards solutions that reduce both energy and time, it can obscure whether an improvement stems from lower average power, shorter runtime, or both. To make this decomposition explicit, the \gls{GSF} proposes the GPS-UP ratio, which reports Speedup (runtime improvement), Powerup (change in average power consumption), and Greenup (overall energy savings combining runtime and power changes) relative to a baseline~\cite{green_software_foundation_gps-up_2021}. This separation supports a clearer interpretation of optimization effects and more transparent decisions under energy and performance constraints.

\begin{table*}[htbp]
\centering
\renewcommand{\arraystretch}{1.05}
\caption{Overview of Energy and Power Efficiency Metrics}
\label{tab:energy-efficiency-metrics}
\begin{tabular}{@{}p{2.5cm}p{6.5cm}p{5cm}p{2cm}@{}}
\toprule
\textbf{Element}         & \textbf{Metric Name}                                                          & \textbf{Formula}                                                                                  & \textbf{References}                                                                  \\ \toprule
\multirow{3}{*}{CPU}               & FLOPS per Watt                                                        & \( \frac{\text{FLOPS}}{\text{Watt}} \)                                                   & \cite{kipp_layered_2012, kipp_applying_2012, beik_green_2012}                                                        \\ \cmidrule(l){2-4}
               & MHz per Watt                                                          & \( \frac{\text{MHz}}{\text{Watt}} \)                                                     & \cite{kipp_layered_2012}                                                        \\ \cmidrule(l){2-4}
               & Power per Utilization                                                & \( \frac{P_{\text{component}}}{U_{\text{component}}} \)                                  & \cite{chen_profiling_2011, beik_green_2012}                  \\ \midrule
\multirow{2}{*}{Storage}           & Capacity per Watt                                                     & \( \frac{\text{Capacity}}{\text{Watt}} \)                                                & \cite{kipp_layered_2012}                                                        \\ \cmidrule(l){2-4}
           & \gls{IOPS} per Watt                                                         & \( \frac{\text{\gls{IOPS}}}{\text{Watt}} \)                                                    & \cite{kipp_layered_2012}                                                        \\ \midrule
Server / Rack     & \gls{SWaP}                                   & \( \frac{\text{Performance}}{\text{Space} \times \text{Power}} \)          & \cite{kipp_layered_2012}                                                        \\ \midrule
\multirow{14}{*}{System of Systems} & \gls{DH-UR}                                & \( \frac{\text{Active Servers}}{\text{Total Deployed Servers}} \)                       & \cite{kipp_layered_2012}                                                        \\ \cmidrule(l){2-4}
 & \gls{DH-UE}                           & \( \frac{\text{Min. Servers for Peak Load}}{\text{Total Deployed Servers}} \)           & \cite{kipp_layered_2012}                                                        \\ \cmidrule(l){2-4}
 & \gls{PUE}                                      & \( \frac{\text{Total Facility Power}}{\text{IT Equipment Power}} \)                     & \cite{kipp_layered_2012, beik_green_2012, kliazovich_greencloud_2012}                                                        \\ \cmidrule(l){2-4}
& \gls{DCIE}                         & \( \frac{\text{IT Equipment Power}}{\text{Total Facility Power}} \)                     & \cite{kipp_layered_2012, kliazovich_greencloud_2012}                                                        \\ \cmidrule(l){2-4}
 & \gls{CPE}                                       & \( \frac{\text{\gls{DCIE}}}{\text{IT Equipment Utilization}} \)                               & \cite{kipp_layered_2012}                                                        \\ \cmidrule(l){2-4}
 & \gls{DC-EEP}             & \( \text{IT-PEW} \times \text{SI-EER} \)                                                 & \cite{kipp_layered_2012}                                                        \\ \cmidrule(l){2-4}
 & \gls{SI-EER}                 & \( \frac{\text{Conditioned Power Out}}{\text{Power In}} \)                              & \cite{kipp_layered_2012}                                                        \\ \cmidrule(l){2-4}
 & \gls{SI-POM}               & \( \frac{\text{Data Centre Power at Meter}}{\text{Total Hardware AC Power}} \)          & \cite{kipp_layered_2012}                                                        \\ \cmidrule(l){2-4}
 & IT Hardware Power Overhead Multiplier (H-POM)                        & \( \frac{\text{Total Hardware Load at Plug}}{\text{Total Compute Load}} \)              & \cite{kipp_layered_2012}                                                        \\ \cmidrule(l){2-4}
 & \gls{DCeP}                               & \( \frac{\text{Work Output (Bytes)}}{\text{Total Energy}} \)                            & \cite{kipp_layered_2012}                                                        \\ \cmidrule(l){2-4}
 & \gls{DCPE}                            & \( \frac{\text{Effective IT Workload}}{\text{Total Facility Power}} \)                  & \cite{kipp_layered_2012}                                                        \\ \cmidrule(l){2-4}
 & Coefficient of Performance (COP Ensemble)                            & \( \frac{\text{Total Heat Dissipation}}{\text{Flow Work} + \text{Thermodynamic Work}} \) & \cite{kipp_layered_2012}                                                        \\ \cmidrule(l){2-4}
 & Carbon Usage Effectiveness (CUE) & \( \text{CUE} = \frac{\text{Total Emissions from Data Center Energy}}{\text{IT Equipment Energy}} \) & \cite{priya_survey_2013} \\ \cmidrule(l){2-4}
 & Green Power Usage Effectiveness (GPUE) & $\text{GPUE} = (G \times \text{PUE})$ \newline with G as the weighted sum of energy sources and their life-cycle carbon intensity & \cite{priya_survey_2013} \\ \midrule
Network           & Bandwidth per Watt                                                    & \( \frac{\text{Bandwidth}}{\text{Watt}} \)                                               & \cite{kipp_layered_2012}                                                        \\ \midrule
\multirow{4}{*}{Software}           & Energy–Delay Product (EDP)                                          & \( E_{\text{task}} \times T_{\text{task}} \)                                              & \cite{green_software_foundation_gps-up_2021} \\ \cmidrule(l){2-4}
           & Speedup                                                             & \( \frac{T_{\text{baseline}}}{T_{\text{optimized}}} \)                                   & \cite{green_software_foundation_gps-up_2021} \\ \cmidrule(l){2-4}
          & Powerup                                                             & \( \frac{P_{\text{baseline}}}{P_{\text{optimized}}} \)                                   & \cite{green_software_foundation_gps-up_2021} \\ \cmidrule(l){2-4}
           & Greenup                                                             & \( \frac{E_{\text{baseline}}}{E_{\text{optimized}}} \)                                   & \cite{green_software_foundation_gps-up_2021} \\
\bottomrule
\end{tabular}
\end{table*}

\paragraph{Environmental Impact Metrics}
\label{ee-measurement-classification-environmental-impact}

Power, energy, and efficiency metrics quantify electricity use and its relation to workload, but they do not capture the broader sustainability implications of \gls{ICT} systems. Environmental impact metrics therefore translate energy use into environmental consequences and, in some cases, extend the assessment beyond operational electricity use to additional life-cycle factors.

In the reviewed literature, this translation is most commonly expressed via carbon-related indicators, including CO$_2$, greenhouse gas emissions, carbon footprint, and \gls{CO2e}~\cite{naqi_ai_nodate,kipp_applying_2012,priya_survey_2013}. Beyond direct carbon conversion, metric systems and reference models are used to account for further impact drivers. Prior work emphasizes that instantaneous power or energy consumption alone is insufficient for a comprehensive sustainability assessment. Instead, the impacts associated with electricity generation, power supply, material use, and emissions should be considered across the software life cycle. 

Kipp et al.~\cite{kipp_green_2011,kipp_applying_2012} exemplify this broader perspective through indices that extend beyond operational power draw. Their consumables index captures the amount of consumables generated during an application's workflow execution. Their electricity consumption index aggregates  total electricity use, including overheads such as cooling and network devices. Their supply chain index accounts for CO$_2$ emissions from transport and logistics associated with the execution of  corresponding services, expressed in CO$_2$ units. In addition, some studies include redevelopment and reconfiguration costs as part of long-term sustainability evaluation~\cite{kipp_layered_2012}.

\subsubsection{Measurement Approaches and Tooling}
\label{ee-measurement-approaches-tooling}
We identified common measurement methodologies and tools across the sustainability metric domains introduced in Section~\ref{ee-measurement-classification}: Section~\ref{ee-measurement-approaches-tooling-energy-consumption} covers approaches and tooling for power and energy consumption, Section~\ref{ee-measurement-approaches-tooling-energy-efficiency} summarizes efficiency indicators derived from these measurements, and Section~\ref{ee-measurement-approaches-tooling-environmental-impact} discusses approaches for environmental impact quantification. In the following, we highlight key approaches and representative tools per domain. A consolidated overview of additional tools and approaches is provided in Table~\ref{tab:overview-pee-ee-ei}.

\paragraph{Approaches and Tooling for Power and Energy Consumption}
\label{ee-measurement-approaches-tooling-energy-consumption}
We identified three overarching approaches for measuring power and energy: external measurement, on-device measurement, and model-based estimation. At the hardware level, all three can be applied. For software elements, power and energy are typically not measured directly but attributed via model-based estimation using underlying hardware measurements and resource-usage data.

\textit{Hardware Level:}
External measurement provides coarse-grained readings at the level of complete devices or subsystems, such as racks or servers. Common tools include \glspl{PDU}, which measure per-outlet power and support remote monitoring~\cite{noauthor_q-rzenamel_nodate,chen_profiling_2011}; standalone power meters such as Watts Up Pro, which are typically placed inline between a device and the power source~\cite{noureddine_review_2013,guldner_development_2024,armstrong_towards_2017}; and \gls{IPMI}, which exposes server-level power information~\cite{jiang_edge_2019}. 

On-device measurement relies on sensors and vendor interfaces to obtain finer-grained readings for specific components. For CPUs and memory, Intel \gls{RAPL} is frequently used to access energy and power data~\cite{jiang_edge_2019,guldner_development_2024,noauthor_q-rzenamel_nodate}. For NVIDIA GPUs, the \texttt{nvidia-smi} toolkit is commonly used to retrieve power readings and, depending on the device and driver support, additional energy-related counters~\cite{guldner_development_2024}.

Model-based approaches estimate power or energy by correlating system activity, for example CPU utilization and I/O operations, with empirical measurements or documented power profiles. 
Models range from simple linear functions to more complex regression-based estimators. Hardware-oriented methods often define component-specific estimation for functions and aggregate them to approximate total system consumption. A common pattern is scaling between idle and maximum power as a function of utilization~\cite{drraghavendra_holistic_2024,priya_survey_2013,daraghmeh_power_2019}. Other work applies domain-specific formulas to capture subsystem contributions, including cooling, storage, computation, and idle operation~\cite{uchechukwu_energy_2014}~\cite{chen_energy_2012}~\cite{singh_estimating_2013}. Advanced estimators incorporate hardware performance counters and OS-level metrics to derive finer-grained models for processors, memory, storage, and network interfaces~\cite{noureddine_review_2013,naqi_ai_nodate}. Supporting tools reported in the literature include PowerTop, Joulemeter, and vMeter~\cite{noureddine_review_2013,priya_survey_2013}.  

\textit{Software Level:}
To estimate the power or energy consumption of software elements, ranging from full software  applications and containers to smaller units such as functions or code snippets, model-based attribution is used. In these approaches, measured or estimated hardware energy is mapped to software entities based on their resource usage over time.

One class of methods applies state-based modeling by categorizing execution into states such as active, idle, or blocked (e.g., waiting on I/O). Each state is associated with an estimated energy cost, and total energy is obtained by aggregating costs over the time spent in each state~\cite{noureddine_review_2013}. 

Another class uses component-based modeling, which decomposes software into logical units such as computation, communication, or infrastructure components. Energy is attributed using system metrics and, where available, hardware performance counters, for example CPU usage, memory accesses, or I/O operations. Variants of this approach are also reported by Noureddine et al.~\cite{noureddine_review_2013}, where application elements are disassembled and modeled separately.

\begin{table*}[htbp]
\centering
\caption{Overview of Tools and Approaches for Power and Energy Estimation, Energy Efficiency and Environmental Impact}
\label{tab:overview-pee-ee-ei}
\begin{tabular}{p{2.5cm}p{3.5cm}p{3cm}p{7cm}}
\toprule
\textbf{Aspect} & \textbf{Representative Approach}                            & \textbf{Target Components}                                      & \textbf{Representative Tools and Methods}                                                                                                                                                                                   \\ \toprule
\multirow{4}{*}{Power / Energy}    
    & External Measurement                         & Racks, Servers                                                  & 
    \glspl{PDU}~\cite{bergen_towards_2014, chen_profiling_2011, guldner_development_2024, armstrong_towards_2017}, Watts Up Pro~\cite{noureddine_review_2013, guldner_development_2024}, \gls{IPMI}~\cite{jiang_edge_2019}                                                                 \\ \cmidrule(l){2-4} 
    & On-Device Measurement                        & CPU, DRAM, GPU                                                  & Intel 
    \gls{RAPL}~\cite{jiang_edge_2019,guldner_development_2024,bergen_towards_2014}, nvidia-smi~\cite{guldner_development_2024}                                                                                                                           \\ \cmidrule(l){2-4} 
    & Model-Based Estimation (HW)                  &CPU, Memory, Network                                    & 
    PowerTop~\cite{noureddine_review_2013}, Joulemeter~\cite{noureddine_review_2013}, vMeter~\cite{priya_survey_2013}                                                                                                  \\ \cmidrule(l){2-4} 
    & \multirow{3}{*}{Model-Based Estimation (SW)} & VMs, Containers,
     Applications               & 
    PowerScope ~\cite{noureddine_review_2013}, pTop ~\cite{noureddine_review_2013}, Jalen ~\cite{noureddine_review_2013}, PowerAPI ~\cite{noureddine_review_2013}, PowerTop ~\cite{noureddine_review_2013}, SERENA~\cite{guldner_development_2024}, Energy Checker ~\cite{noureddine_review_2013}, GSMP~\cite{guldner_development_2024}, SAF Toolkit and Green Lab~\cite{guldner_development_2024}, Green Metrics Tool~\cite{guldner_development_2024}, SPECPower XGBoost Model~\cite{guldner_development_2024}, SDIA Framework and Linear Approximation~\cite{guldner_development_2024}, Oeko-Institute Software Footprint Tool~\cite{guldner_development_2024}, Kepler~\cite{apiumhub_18_2024}, PowerNode~\cite{chen_energy_2012}, EdgeCloudSim~\cite{daraghmeh_power_2019}, Energy Metering Tool~\cite{singh_estimating_2013}, GreenDroid~\cite{apiumhub_18_2024}
    \\ \cmidrule(l){3-4} 
    &                                               & Code snippets, routines, functions & 
    SERENA~\cite{guldner_development_2024}, GSMP~\cite{guldner_development_2024}, SAF Toolkit and Green Lab~\cite{guldner_development_2024}, SDIA Framework and Linear Approximation~\cite{guldner_development_2024}, Oeko-Institut’s Software Footprint Tool~\cite{guldner_development_2024}, CodeCarbon~\cite{CodeCarbon}, Carbontracker ~\cite{apiumhub_18_2024}, JoularJX~\cite{apiumhub_18_2024}, GCPM~\cite{nasrin_promoting_2024}, Eco CI~\cite{noauthor_green-coding-solutionseco-ci-energy-estimation_nodate}                                            \\ \midrule

\multirow{2}{*}{Energy Efficiency}         
    & \multirow{2}{*}{Model-Based Estimation (SW)}                  & VMs, Containers, Applications               & 
    SAF Toolkit and Green Lab~\cite{guldner_development_2024}, Green Metrics Tool~\cite{guldner_development_2024}, Oeko-Institut’s Software Footprint Tool~\cite{guldner_development_2024}, Customer Cloud Carbon Footprint ~\cite{apiumhub_18_2024}, AWS Green Cost Explorer~\cite{GreenCostExplorer}, Emissions Impact Dashboard ~\cite{apiumhub_18_2024}, Cloud Carbon Footprint ~\cite{apiumhub_18_2024}, Google Carbon Footprint ~\cite{apiumhub_18_2024}
    
    \\ \cmidrule(l){3-4} 
                  &   & Code snippets, routines, functions & 
    SAF Toolkit and Green Lab~\cite{guldner_development_2024}, Oeko-Institut’s Software Footprint Tool~\cite{guldner_development_2024}, ENRICO~\cite{lange_energy_2025}, EcoCode ~\cite{lange_energy_2025}, Ec0lint ~\cite{lange_energy_2025}                    \\ \midrule

\multirow{2}{*}{Environmental Impact}    
    & \multirow{2}{*}{Model-Based Estimation (SW)}                  & VMs, Containers, Applications, Web Pages               & 
    Green Metrics Tool~\cite{guldner_development_2024}, SAF Toolkit and Green Lab~\cite{guldner_development_2024}, Oeko-Institut’s Software Footprint Tool~\cite{guldner_development_2024}, Website Carbon Calculator~\cite{WebsiteCarbonCalculator}, Customer Carbon Footprint Tool ~\cite{CarbonFootprintTool}, AWS Green Cost Explorer~\cite{GreenCostExplorer}, Emissions Impact Dashboard ~\cite{apiumhub_18_2024}, Cloud Carbon Footprint ~\cite{apiumhub_18_2024}, Google Carbon Footprint ~\cite{apiumhub_18_2024}, Carbontracker ~\cite{apiumhub_18_2024}, Fruggr ~\cite{apiumhub_18_2024}
    
    \\ \cmidrule(l){3-4} 
    &                  & Code snippets, routines, functions & 
    SAF Toolkit and Green Lab~\cite{guldner_development_2024}, Oeko-Institut’s Software Footprint Tool~\cite{guldner_development_2024}, CodeCarbon~\cite{CodeCarbon}, EcoCode ~\cite{lange_energy_2025}, Ec0lint ~\cite{lange_energy_2025}
    
    \\ \bottomrule
\end{tabular}
\vspace{2pt}
\begin{flushleft}
\footnotesize (HW) indicates estimation based on hardware components; (SW) indicates estimation based on software components.
\end{flushleft}
\end{table*}

\paragraph{Approaches and Tooling for Energy Efficiency}
\label{ee-measurement-approaches-tooling-energy-efficiency}
Approaches and tooling for energy efficiency largely build on the power and energy measurement mechanisms described in Section~\ref{ee-measurement-approaches-tooling-energy-consumption}. Instead of  introducing new measurement techniques, most works derive efficiency indicators from existing power and energy data by normalizing consumption to a defined functional output or workload. Typical examples include transactions per kWh, GreenUp, or power per utilization, which support comparison across configurations, deployments, or code revisions under controlled workloads. Accordingly, the literature reports few tools dedicated primarily to energy efficiency rather than power/energy metering.

For example, the Customer Carbon Footprint Tool and Green Cost Explorer provide combined views of cost and sustainability-related indicators for cloud workloads~\cite{apiumhub_18_2024}. Such dashboards help  users identify energy and carbon hotspots and relate them to cost drivers, supporting trade-off analyses across deployments and regions.

Moreover, GCPM (GreenCode-PowerMeter) is a model-based power estimation framework integrated into VS Code. It allows developers to estimate and compare the energy consumption of code directly within the  development workflow, without requiring dedicated measurement hardware or administrative privileges~\cite{nasrin_promoting_2024}. This supports rapid feedback on the sustainability implications of code changes and can guide energy-aware refactoring and optimization decisions.

\paragraph{Approaches and Tooling for Environmental Impact}
\label{ee-measurement-approaches-tooling-environmental-impact}
Beyond low-level power and energy estimation, standardized approaches for quantifying the environmental impact of software workloads are gaining importance. These approaches typically translate measured or estimated energy use into impact indicators (most commonly \gls{CO2e}) and, where possible, extend the perspective to life-cycle effects. 

The GREENSOFT Model provides a reference model for sustainable software engineering with an explicit life-cycle perspective, spanning requirements, development, operation, and decommissioning. It considers environmental aspects such as energy use over the life cycle and resource utilization and recommends process-level practices, for example regular sustainability reviews and clearly assigned responsibilities among developers, system administrators and end users~\cite{naumann_greensoft_2011}.

The \gls{SCI} specification offers a unified framework for attributing carbon emissions to software workloads by combining operational energy consumption \(E\), grid carbon intensity \(I\), embodied emissions \(M\), and functional output \(R\) into a single intensity metric~\cite{tarara_automated_2023}. This supports benchmarking and comparability across use cases and infrastructures, provided that the functional unit and system boundaries are defined consistently.

To support adoption, the \gls{SCI} Guidance project provides detailed methodologies for estimating the parameters of the \gls{SCI} equation and for selecting appropriate functional units~\cite{tarara_automated_2023}. In addition, the Green Software Patterns repository contributes a curated catalog of patterns that target energy and carbon reduction across recurring architectural concerns~\cite{green_software_foundation_measuring_2023}. Together, these resources help practitioners integrate environmental impact considerations into design and implementation decisions early in the software development life cycle.

\subsection{Software- and System-Level Optimization of Power Consumption and Energy Efficiency in ICT Systems}
\label{ee-optimization}

The literature proposes a broad range of methods and techniques to improve energy efficiency and reduce power consumption in \gls{ICT} systems, with a strong emphasis on \glspl{DCCCE}. To address RQ3, this section focuses on software and system strategies in these environments, excluding measures primarily concerned with hardware design or physical facility engineering. This scope aligns with the software-centric perspective of this \gls{MLR} and targets interventions that can be applied through software, orchestration, and architectural decisions. Such approaches are widely regarded as offering substantial energy-saving potential in practice~\cite{khattar_toward_2019,kanso_review_2024}.

To structure the analysis, we synthesize 25 sources into four thematic groups that reflect common optimization directions in the literature: Section~\ref{ee-optimization-dynamic-power} discusses Dynamic Power and Resource Management, including runtime adaptations such as task placement, workload distribution, and voltage and frequency control in response to changing demands; Section~\ref{ee-optimization-vm-management} covers VM Management and Server Consolidation Methods, focusing on placement and migration strategies to reduce idle capacity; Section~\ref{ee-optimization-energy-aware-design} summarizes Energy-Aware Software and System Design, emphasizing architectural and development-time decision; and Section~\ref{ee-optimization-system-wide-efforts} reviews System-Wide and Macro-Level Optimization Efforts that coordinate energy management across layers and components.

\subsubsection{Dynamic Power and Resource Management}
\label{ee-optimization-dynamic-power}
Dynamic power and resource management is a core line of work in energy-efficient \gls{ICT} system design. It focuses on runtime adaptation of resource allocation and power states in response to changing demand and operational context. Typical mechanisms include shifting workloads, consolidating tasks, and adjusting voltage/frequency to manage the energy performance trade-off while meeting service requirements~\cite{khattar_toward_2019,katal_energy_2023}.

Khattar et al.~\cite{khattar_toward_2019} survey heuristics-based, scheduling-based, and Dynamic Power Management (DPM) techniques in \glspl{DCCCE}, highlighting mechanisms such as \gls{DVFS}, energy-aware task scheduling, and adaptive workload balancing. Priya et al.~\cite{priya_survey_2013} classify related techniques along the cloud stack, distinguishing interventions at the physical, middleware, and application layers, and emphasize that energy modeling is a prerequisite for effective scheduling policies. Complementing these surveys, Katal et al.~\cite{katal_energy_2023} provide a detailed discussion of software-level optimization mechanisms, including application-aware workload classification, load balancing across containers in microservices environments, and controls for resource over-provisioning. They stress that energy-saving techniques should be aligned with \glspl{SLA} to ensure that sustainability improvements do not come at the expense of service quality.

A specialized example is iSwitch~\cite{li_iswitch_2012}, a lightweight, application-independent, supply-aware energy management framework for \glspl{DCCCE} powered by intermittent renewable energy sources. By coordinating load migration and server power states with fluctuations in renewable availability, iSwitch aims to increase renewable energy utilization while reducing network congestion and job latency.

Hintemann and Hinterholzer~\cite{hintemann_energy_nodate} contextualize these approaches by highlighting broader trends in data center energy use. They report that although energy efficiency per unit of computing has improved, total electricity consumption continues to increase due to rising data traffic, storage demand, and emerging workloads such as AI and machine learning. 
Johnston~\cite{johnston_improving_2023} similarly argues for intelligent orchestration and management to address inefficiencies from static provisioning and advocates for embedding green SE principles into system design to support holistic energy-aware architectures.

\subsubsection{VM Management and Server Consolidation Methods}
\label{ee-optimization-vm-management}

Energy-efficient \gls{VM} management and server consolidation are central to reducing power consumption in \glspl{DCCCE}, because they target energy waste from underutilized or idling servers. The core idea is to adapt \gls{VM} placement dynamically to operational requirements, balance load across \glspl{PM}, and transition idle \glspl{PM} into low-power states or power them off. By reducing the number of active \glspl{PM} while preserving required capacity, substantial energy savings can be achieved~\cite{hosseini_shirvani_survey_2020}.

A review of \gls{VM} consolidation approaches identifies live migration as a key enabling mechanism, allowing \glspl{VM} to be moved without service interruption in response to workload fluctuations and changing infrastructure conditions~\cite{ahmad_virtual_2015}. Migration, however, is resource-intensive and introduces a trade-off between energy savings, migration overhead, and the risk of \gls{SLA} violations. To improve this trade-off, several works propose heuristics and multi-objective optimization strategies. Azizi et al.~\cite{azizi_energy-efficient_2020} introduce a placement approach that uses a reward and penalty model to reduce the number of active \glspl{PM}. Yakhchi et al.~\cite{yakhchi_proposing_2015} apply a Cuckoo Optimization Algorithm combined with a minimum migration time policy to reduce \gls{SLA} violations and the number of migrations.

Beyond \gls{VM}-centric consolidation, container-based approaches have gained relevance for finer-grained resource management and are often reported to incur lower overhead than \gls{VM}-based solutions. Gholipour et al.~\cite{gholipour_novel_2020} propose a joint \gls{VM} and container consolidation framework that applies a multi-criteria decision model to reduce energy consumption while minimizing \gls{SLA} violations and migration frequency. Their results indicate that  combining \glspl{VM} and containers can outperform isolated consolidation strategies by leveraging the lightweight reconfiguration capabilities of containers.

At a broader system level, workload correlation has been used to improve consolidation decisions by accounting for temporal patterns in resource demand. Power-NetS~\cite{zheng_joint_2014} jointly optimizes server and network energy consumption by analyzing correlations between server workloads and network traffic. By avoiding placements where correlated workloads peak simultaneously, it aims to prevent coinciding overload and reduce the risk of congestion and network delay.

Empirical studies further emphasize the importance of real-time monitoring and operational integration. Atiewi et al.~\cite{atiewi_impact_2018} and Yang and Wan~\cite{yang_implementation_2020} highlight the role of dynamic migration in practice. Yang and Wan~\cite{yang_implementation_2020} demonstrate feasibility in an OpenStack-based setup that monitors power consumption and balances load in real time to optimize energy use.

Finally, taxonomies and surveys provide additional structure for the evolving design space of placement and migration algorithms. Choudhary et al.~\cite{choudhary_critical_2016} classify \gls{VM} placement approaches by objective functions, including energy, \gls{SLA}, and resource constraints, and highlight the use of dynamic thresholds and predictive models. Hsu et al.~\cite{hsu_optimizing_2014} complement this line of work by showing that energy-aware task consolidation, e.g., by limiting CPU utilization via strategic grouping across virtual clusters, can yield measurable reductions in energy use.

\subsubsection{Energy-Aware Software and System Design}
\label{ee-optimization-energy-aware-design}

Energy-aware software and system design integrates energy efficiency considerations into software development and system architecture, aiming to address sustainability already at design time. The literature suggests that treating energy awareness as a first-class design concern can yield substantial savings, particularly when aligned with system-level objectives and operational constraints in \glspl{DCCCE}~\cite{sofia_implementation_2015, manimegalai_energy_2024}.

Sofia and Kumar~\cite{sofia_implementation_2015} argue that energy awareness should be a core design principle for sustainable and cost-effective cloud systems. In the context of "Green cloud", they propose a framework that combines mechanisms such as dynamic power management, \gls{VM} management and task scheduling algorithms within a simulated \gls{DCCCE}, illustrating how energy-aware design choices can be embedded across system layers.

Bergen et al.~\cite{bergen_towards_2014} address the need to align software behavior with energy efficiency goals by proposing a power profiling method that classifies processes by CPU and memory 
demand based on measurements from server racks. Their results indicate that even for constant  total compute work, distributing workloads across different numbers of cores can lead to distinct  energy consumption patterns, highlighting that software-driven workload distribution affects overall power use and efficiency.

Luo et al.~\cite{luo_resource_2012} propose an energy-efficient resource scheduling algorithm tailored to \glspl{DCCCE} that adapts the operating states of components such as CPU, memory and network interfaces based on workload class (e.g., compute-~, storage-~, or I/O-intensive). They report large gains for workloads that would otherwise not fully utilize provisioned resources.

From a developer-centric perspective, Manimegalai et al.~\cite{manimegalai_energy_2024} emphasize that implementation decisions directly influence how computational resources are used. They report measurable savings from energy-aware choices in data structures, control flow, and memory access patterns, supported by an empirical comparison of a conventional implementation and an energy-optimized variant. Complementing this, Meinhardt~\cite{meinhardt_software_2023} presents a catalog of design tactics and trade-offs for integrating  energy efficiency into early-stage design and implementation, arguing that teams should combine architecture-level decisions with empirical energy data to systematically incorporate sustainability objectives into SE practice.

\subsubsection{System-Wide and Macro-Level Optimization Efforts}
\label{ee-optimization-system-wide-efforts}
Several studies emphasize cross-layer, system-wide optimization approaches that span infrastructure, software, and operations in \glspl{DCCCE}. The shared rationale is that substantial energy savings require coordinated interventions across multiple layers rather than isolated 
optimizations~\cite{park_energy_2017,kanso_review_2024,uchechukwu_energy_2014}.

Park et al.~\cite{park_energy_2017} analyze how cloud computing and IT investment influence energy efficiency. Their results indicate that cloud-based services, modeled as firms outsourcing data processing and hosting to cloud providers, can improve client-side energy efficiency. A key driver is that centralized cloud data centers can benefit from of economies of scale, resource pooling, and more efficient infrastructure. The study also suggests that additional internal IT investments during cloud adoption can further amplify these efficiency gains.

Kanso et al.~\cite{kanso_review_2024} argue that architectural energy optimization strategies should be connected across layers. In this view, energy efficiency should not be addressed in isolation at individual components or system levels, but through integrated, system-wide coordination. Focusing on cyber-physical systems, they identify a lack of holistic and autonomous optimization approaches and propose an energy-oriented reference architecture that monitors, analyzes, and controls energy consumption across layers.

Uchechukwu et al.~\cite{uchechukwu_energy_2014} contribute a holistic modeling approach for analyzing and optimizing energy consumption in cloud computing data centers. They introduce a layered energy model that distinguishes fixed components (e.g., idle servers, infrastructure overhead) from dynamic components (e.g., active computation, storage, cooling). This distinction enables fine-grained system-level reasoning. By simulating configurations such as improved virtualization, containerization and cooling control they show that coordinated strategies can yield substantial savings. Their findings reinforce the need for integrated architectural planning that covers the full stack of physical and virtual infrastructure, including workload distribution, task scheduling, and thermal management.

Uddin and Rahman~\cite{uddin_energy_2012} operationalize such energy-aware simulations through a phased implementation framework. Their approach segments data centers into measurable units, applies green metrics such as power usage effectiveness, data center efficiency, and data center productivity, and integrates virtualization to simulate and benchmark system-wide energy savings under realistic workloads. They emphasize that phased, metrics-driven approaches support systematic evaluations while enabling continuous benchmarking and improvement across the infrastructure.

\begin{table*}[htbp]
\caption{Overview of Optimization Approaches, Methods, and Techniques}
\label{tab:energy-optimization}
\centering
\begin{tabular}{@{}p{2cm} p{2.9cm} p{4.2cm} p{4.7cm} p{2cm}@{}}
\toprule
\textbf{Optimization Approach} & \textbf{Objective} & \textbf{Main Methods} & \textbf{Key Insights} & \textbf{References} \\
\toprule
Dynamic Resource \& Power Management & Adjust energy consumption dynamically based on real-time demand. & 
\begin{minipage}[t]{\linewidth}
\begin{itemize}[leftmargin=*]
  \item DVFS
  \item Dynamic scheduling
  \item Load balancing
  \item SLA-aware control
  \item Supply-aware frameworks
\end{itemize}\end{minipage} &
\begin{minipage}[t]{\linewidth}
\begin{itemize}[leftmargin=*]
  \item Real-time adaptation can maintain performance while saving energy.
  \item Integration with renewable sources enhances sustainability.
\end{itemize}\end{minipage} & 
\cite{khattar_toward_2019,priya_survey_2013,katal_energy_2023,li_iswitch_2012,hintemann_energy_nodate,johnston_improving_2023} \\
\midrule
VM \& Server Consolidation & Minimize active servers and optimize \gls{VM}/container placement. & 
\begin{minipage}[t]{\linewidth}
\begin{itemize}[leftmargin=*]
  \item VM/container consolidation
  \item Live migration
  \item Heuristic and predictive placement
  \item Workload correlation-aware scheduling
\end{itemize}\end{minipage} & 
\begin{minipage}[t]{\linewidth}
\begin{itemize}[leftmargin=*]
  \item Consolidation greatly reduces energy use, but migration cost and SLA adherence are critical.
  \item Containers offer better flexibility and efficiency than \glspl{VM}.
\end{itemize}\end{minipage} & 
\cite{yakhchi_proposing_2015,hosseini_shirvani_survey_2020,ahmad_virtual_2015,zheng_joint_2014,gholipour_novel_2020,azizi_energy-efficient_2020,atiewi_impact_2018,yang_implementation_2020,choudhary_critical_2016,hsu_optimizing_2014,ahmad_virtual_2015,ye_energy-efficient_2017} \\
\midrule
Energy-Aware Software/System Design & Integrate energy considerations into software/system architecture. &
\begin{minipage}[t]{\linewidth}
\begin{itemize}[leftmargin=*]
  \item Energy-aware scheduling
  \item Workload profiling
  \item Green coding patterns
  \item Energy-aware software design
\end{itemize}\end{minipage} & 
\begin{minipage}[t]{\linewidth}
\begin{itemize}[leftmargin=*]
  \item Developer design choices significantly affect power use.
  \item Early integration of energy awareness yields sustainable systems.
\end{itemize}\end{minipage} & 
\cite{sofia_implementation_2015,manimegalai_energy_2024,bergen_towards_2014,luo_resource_2012,meinhardt_software_2023} \\
\midrule
System-Wide \& Macro-Level Optimization & Coordinate energy optimization across system/infrastructure layers. & 
\begin{minipage}[t]{\linewidth}
\begin{itemize}[leftmargin=*]
  \item Layered energy models
  \item Green metrics
  \item Cross-layer integration
  \item Energy-aware cloud architectures
\end{itemize}\end{minipage} & 
\begin{minipage}[t]{\linewidth}
\begin{itemize}[leftmargin=*]
  \item Holistic, metric-driven, phased approaches enable large-scale improvements.
  \item Cloud infrastructures often yield systemic efficiency gains.
\end{itemize}\end{minipage} & 
\cite{park_energy_2017,kanso_review_2024,uchechukwu_energy_2014,uddin_energy_2012} \\
\bottomrule
\end{tabular}
\end{table*}

\subsection{Code Generation and Transformation for Resource-Efficient Software in ICT Systems}
\label{ee-code-optimization}

The optimization strategies reviewed above operate above the code layer, at the level of scheduling, placement, resource management, and system design. A complementary and rapidly emerging class of techniques instead targets the code layer directly, using \glspl{LLM} to generate or transform code for improved resource efficiency. We review this layer next.

Interest in AI-assisted software development has intensified, and with it the attention to \glspl{LLM} for code generation and transformation. Recent work investigates whether these models can produce code that improves runtime performance, reduces memory usage, and, where measured, lowers energy consumption. In this section, we examine how \glspl{LLM} support performance- and energy-aware development by automating code-level tasks such as generation, refactoring, and efficiency-oriented transformation.

To address RQ3, we analyze how resource efficiency is evaluated in both academic and practitioner literature. While some studies directly measure energy (e.g., via profilers or hardware interfaces), many evaluate efficiency indirectly through performance proxies such as  execution time and memory usage. We include both types of evidence, since performance improvement can translate into energy savings depending on workload and platform characteristics.

To keep the scope focused, we consider only work in which \glspl{LLM} are used to generate code or to transform existing code with the explicit goal of efficiency improvements. We exclude studies that primarily analyze the energy or computational footprint of the \glspl{LLM} themselves.

\newpage
To structure our analysis, we categorize 21 recent studies across seven dimensions: 
(i) \emph{task formulation} (generation vs. transformation), (ii) \emph{nature of contribution} (application, method, benchmark, or survey), (iii) \emph{targeted programming languages}, (iv) \emph{prompting strategies}, (v) \emph{benchmarks and datasets}, (vi) \emph{evaluation metrics}, and (vii) \emph{training setup} (when model training is involved). 
This structure provides a consolidated view of current practice in efficiency-oriented, 
\gls{LLM}-assisted code generation and optimization. 

\subsubsection{LLM Task Formulation}
\label{ee-code-optimization-task-formulation}

Across the reviewed literature, we identify two primary task formulations for applying \glspl{LLM} to code development: \emph{code generation} and \emph{code transformation}. In the code generation setting, \glspl{LLM} are prompted to synthesize complete implementations from natural language descriptions or problem statements. This setting is widely used in studies that evaluate the performance, and in some cases the energy efficiency, of automatically produced code.

Eleven reviewed papers adopt the code generation setting. They include empirical evaluations~\cite{liu_evaluating_2024,shehab_evaluating_2024,vartziotis_learn_2024,tuttle_can_2024,cappendijk_generating_2024,cursaru_controlled_2024}, benchmark and framework proposals~\cite{qiu_how_2025,huang_effibench_2024,huang_effi-code_2024,liu_evaluating_2024}, and methodology-oriented work~\cite{huang_effi-code_2024}. These studies prompt \glspl{LLM} to solve tasks from benchmarks such as HumanEval~\cite{chen_evaluating_2021}, LeetCode~\cite{noauthor_leetcode_nodate}, and EffiBench ~\cite{huang_effibench_2024}, and then analyze runtime, energy consumption, and/or memory usage of the generated solutions.

In contrast, the code transformation setting aims to improve the efficiency of existing code. Here, the \gls{LLM} is instructed to modify a given implementation to reduce runtime or energy consumption while preserving correctness. This formulation appears in three papers. Duan et al.~\cite{duan_leveraging_2023} leverage \gls{RL} and reward models to guide optimization, Rosas et al.~\cite{rosas_should_2024} compare LLM-optimized code against compiler-optimized outputs. Peng et al.~\cite{peng_large_2024} apply profiling-guided prompting and iterative refinement for energy-aware optimization. These studies are commonly framed in the context of software maintenance.

Two additional papers are surveys and therefore do not fit the task-based categorization. Gong et al.~\cite{gong_language_2025} survey LLM-based code optimization, and Wang et al.~\cite{wang_enhancing_2025} focus on \gls{RL} for code generation.

\subsubsection{Nature of Contribution}
\label{ee-code-optimization-nature-contribution}

In addition to task formulation, we categorized the reviewed studies based on the nature of their contributions. Each work is assigned to one of the following categories: application, method, benchmark, or survey.

Several studies apply existing \glspl{LLM} to code-related tasks. These often benchmark different models or prompt strategies on programming tasks~\cite{liu_evaluating_2024,shehab_evaluating_2024,vartziotis_learn_2024,tuttle_can_2024,cappendijk_generating_2024,cursaru_controlled_2024,rosas_should_2024,huang_effibench_2024, qiu_how_2025}. While they involve empirical evaluation, their contribution lies in applying \glspl{LLM} to practical use cases and analyzing outcomes under various prompting or optimization setups.

Three papers introduce new methods for enhancing LLM performance. Huang et al.~\cite{huang_effi-code_2024} propose a self-improving code generator leveraging runtime profiling for iterative refinement. This work complements approaches proposed by Duan et~al.~\cite{duan_leveraging_2023} and Peng et~al.~\cite{peng_large_2024}, as outlined in Section~\ref{ee-code-optimization-task-formulation}.

Several studies contribute benchmark frameworks. Liu et al.~\cite{liu_evaluating_2024} present EvalPerf, a benchmark and evaluation framework for performance-critical code generation. Huang et al.~\cite{huang_effibench_2024} develop EffiBench, a benchmark for assessing the efficiency of code generated by LLMs on algorithmic tasks. Additionally, Huang et al.~\cite{huang_effi-code_2024} propose EffiCode, a dataset for fine-tuning \glspl{LLM} to improve both correctness and efficiency. Qiu et al.~\cite{qiu_how_2025} introduce ENAMEL, a benchmark suite that evaluates the efficiency of \gls{LLM}-generated code using a metric called \textit{eff@k}. Finally, Rosas et al.~\cite{rosas_should_2024} introduce the PCB Suite, a benchmark comprising 28 tasks, each targeting one of six challenging scenarios commonly encountered by optimizing compilers.

Table~\ref{tab:contribution-task} summarizes the reviewed studies by contribution type and task formulation. It shows whether each work focuses on code generation or transformation and highlights its primary research role.

\begin{table}[htbp]
\centering
\caption{Grouped Studies by Nature of Contribution and Task Formulation}
\label{tab:contribution-task}
\begin{tabular}{lll}
\toprule
\textbf{Nature of Contribution} & \textbf{Task Formulation} & \textbf{References} \\
\toprule

\multirow{2}{*}{Application}
& Code Generation 
& \cite{shehab_evaluating_2024,vartziotis_learn_2024,tuttle_can_2024,cappendijk_generating_2024,cursaru_controlled_2024,liu_evaluating_2024, qiu_how_2025, huang_effibench_2024} \\ \cmidrule(l){2-3}
& Code Transformation 
& \cite{rosas_should_2024} \\

\midrule
\multirow{2}{*}{Method}
& Code Generation 
& \cite{huang_effi-code_2024} \\ \cmidrule(l){2-3}
& Code Transformation 
& \cite{duan_leveraging_2023,peng_large_2024} \\

\midrule
\multirow{1}{*}{Benchmark}
& Code Generation 
& \cite{huang_effibench_2024,qiu_how_2025,liu_evaluating_2024,huang_effi-code_2024} \\ \cmidrule(l){2-3}
& Code Transformation 
& \cite{rosas_should_2024} \\

\midrule
Survey 
& N/A 
& \cite{gong_language_2025,wang_enhancing_2025} \\

\bottomrule
\end{tabular}
\end{table}

\subsubsection{Programming Language}
\label{ee-code-optimization-programming-language}
The choice of programming language plays a key role in how LLMs are applied for code generation and optimization tasks, especially in terms of tooling, benchmarks, and profiling support.

Table~\ref{tab:pls} presents the distribution of targeted programming languages across the reviewed studies. Python is by far the most commonly used programming language. C++ appears in three studies. The remaining languages (Java, JavaScript, and C) are each used in only one study. Some studies cover multiple programming languages and are therefore counted in more than one category. Overall, the distribution highlights the dominance of Python in the reviewed LLM-related tasks.

\begin{table}[htbp]
\centering
\caption{Targeted Programming Languages}
\label{tab:pls}
\begin{tabular}{lcl}
\toprule
\textbf{Language} & \textbf{\# of References} & \textbf{References} \\
\toprule
Python & 9 & \cite{tuttle_can_2024,vartziotis_learn_2024,huang_effi-code_2024,huang_effibench_2024,liu_evaluating_2024,qiu_how_2025,duan_leveraging_2023,cappendijk_generating_2024,cursaru_controlled_2024} \\
\midrule
C++ & 3 & \cite{peng_large_2024,cursaru_controlled_2024,shehab_evaluating_2024} \\
\midrule
Java & 1 & \cite{shehab_evaluating_2024} \\
\midrule
JavaScript & 1 & \cite{cursaru_controlled_2024} \\
\midrule
C & 1 & \cite{rosas_should_2024} \\
\bottomrule
\end{tabular}
\end{table}

\subsubsection{Inference Prompting Strategies}
\label{ee-code-optimization-prompting-strategies}

LLM-based code generation and optimization have explored various prompting strategies. In this subsection, these strategies are characterized by the degree of explicit guidance they provide to the model and by whether they support iterative refinement. We categorize these prompting strategies into five main types: (i) \emph{Basic Instruction}, (ii) \emph{Objective-Driven Instruction}, (iii) \emph{Guided Instruction}, (iv) \emph{\gls{CoT}}, and (v) \emph{\gls{ICL}} (one-shot/few-shot). An overview of these categories is presented in Table~\ref{tab:prompts}. Training-related prompting strategies are discussed in detail in Section~\ref{ee-code-optimization-training-setup}.

\emph{Basic Instruction} prompts represent the most straightforward approach, where the model is asked to solve a programming task or generate code without any explicit hints. These prompts typically lack metric-awareness or detailed guidance and are commonly used to establish a performance baseline. Shehab et al.~\cite{shehab_evaluating_2024} use a general prompt template such as "Write a [language] implementation to solve the following programming task". Similarly, Tuttle et al.~\cite{tuttle_can_2024} introduce a baseline prompt "Write a Python program to solve this problem." as part of a comparative framework to study prompt engineering effects. In the same vein, several studies adopt basic instruction prompts that request code generation without any explicit optimization target or performance guidance~\cite{cursaru_controlled_2024,qiu_how_2025,huang_effi-code_2024, liu_evaluating_2024, duan_leveraging_2023}.

\emph{Objective-driven instruction} prompts explicitly embed performance objectives such as improving runtime, reducing memory usage, or minimizing energy consumption into the prompt itself. These can be formulated with direct suffixes like “Give me a runtime-optimized solution for this problem” or “Optimize this code for memory.” Several studies apply such goal-aware prompting~\cite{vartziotis_learn_2024,liu_evaluating_2024,rosas_should_2024,cappendijk_generating_2024,cursaru_controlled_2024,tuttle_can_2024}. In two studies, this strategy is extended with iterative interactions. For example, Peng et al.~\cite{peng_large_2024} introduce Energy-Aware Prompting which combines optimization objectives with profiling-based feedback. 
Qiu et al.~\cite{qiu_how_2025} introduce a refinement method called Self-Refine, in which the model is first asked to "Give a potential strategy to improve the efficiency of the code." followed by a prompt to "Give the optimized version of the same code with the strategy mentioned above." We classify this approach as a partial iteration of objective-driven prompting, as it involves a single refinement step based on an explicit performance goal, but does not include a continuous feedback loop.

\emph{Guided instruction} prompts provide specific suggestions on how to achieve a given task, offering more direction compared to basic or objective-driven instructions. Examples include statements such as "Use a for loop instead of a while loop" or "Consider using dynamic programming." Such guided prompting has been applied in several studies~\cite{rosas_should_2024, cappendijk_generating_2024, qiu_how_2025}. For instance, Peng et al.~\cite{peng_large_2024} incorporate guided refinements within an iterative loop, while Qiu et al.~\cite{qiu_how_2025} evaluate the model's ability to generate efficient algorithms under direct guidance. Extending the Self-Refine prompting strategy discussed earlier, they reveal part of the solution through the prompt such as "Complete the following function with a dynamic programming over digits."

\emph{\gls{CoT}} prompting introduces step-by-step reasoning into the code generation process. \gls{CoT} instructions such as "Think step by step" are employed in~\cite{liu_evaluating_2024, rosas_should_2024} to encourage intermediate reasoning before producing final code outputs. Tuttle et al.~\cite{tuttle_can_2024} implement a two-step prompting strategy, where an initial basic prompt is followed by a refinement prompt such as “Now make this code more energy efficient.” We classify this approach as a form of partial iteration, as it involves a single manual refinement rather than a full multi-turn feedback loop.

\emph{\gls{ICL}} strategies rely on one-shot or few-shot examples to guide generation. For example, Huang et al.~\cite{huang_effibench_2024} use a static prompting setup that includes detailed task descriptions, annotated input–output pairs, function signatures, and accompanying rationales. In contrast, Peng et al.~\cite{peng_large_2024} use a dynamic one-shot setup enriched with CoT rationales and self-consistency, where prompts are refined iteratively based on performance, guiding the model to explore multiple optimization strategies before selecting an efficient solution.

Several studies combine elements from different prompting strategies to enhance code generation and transformation. For example, a single prompt may include both objective-driven instruction and guided hints, or embed \gls{CoT} reasoning within a few-shot example. The literature spans a spectrum of approaches, ranging from static, single-turn prompts to multi-stage prompting schemes that incorporate feedback and explicitly structure the optimization intent.

\begin{table}[htbp]
\centering
\caption{Prompting Strategies in Inference}
\label{tab:prompts}
\begin{tabular}{@{}p{3cm}p{2cm}p{3cm}@{}}
\toprule
\textbf{Prompting Strategy} & \textbf{Iterative Optimization} & \textbf{References} \\
\toprule
\multirow[t]{1}{3 cm}{Basic Instruction Prompt} 
  & No 
  & \cite{liu_evaluating_2024,shehab_evaluating_2024,tuttle_can_2024,cursaru_controlled_2024,qiu_how_2025,huang_effi-code_2024,duan_leveraging_2023} \\
\midrule
\multirow{2}{3 cm}{Objective-Driven Instruction Prompt} 
  & No 
  & \cite{vartziotis_learn_2024,liu_evaluating_2024,rosas_should_2024,cappendijk_generating_2024,cursaru_controlled_2024,tuttle_can_2024} \\ \cmidrule(l){2-3}
  & Partial 
  & \cite{qiu_how_2025} \\ \cmidrule(l){2-3}
& Yes 
  & \cite{peng_large_2024} \\
\midrule
\multirow{2}{3 cm}{Guided Instruction Prompt} 
  & No 
  & \cite{rosas_should_2024,cappendijk_generating_2024} 
   \\ \cmidrule(l){2-3}
  & Partial 
  & \cite{qiu_how_2025} \\ \cmidrule(l){2-3}
& Yes 
  & \cite{peng_large_2024} \\
\midrule
\multirow{3}{3 cm}{Chain-of-Thought Prompt} 
  & No 
  & \cite{liu_evaluating_2024,rosas_should_2024} \\ \cmidrule(l){2-3}
& Partial 
  & \cite{tuttle_can_2024} \\ \cmidrule(l){2-3}
& Yes 
  & \cite{peng_large_2024} \\
\midrule
\multirow{2}{3 cm}{In-Context Learning (1-shot/few-shot)} 
  & No 
  & \cite{huang_effibench_2024} \\ \cmidrule(l){2-3}
& Yes 
  & \cite{peng_large_2024} \\
\bottomrule
\end{tabular}
\end{table}

\subsubsection{Benchmarks}
\label{ee-code-optimization-benchmarks}
Several studies evaluate the effectiveness of LLM-generated code using diverse benchmark types, as summarized in Table~\ref{tab:benchmarks}. These benchmarks differ in their primary objectives ranging from functional and correctness checks to detailed measurements of performance. We categorize them into four groups: (i) \emph{algorithmic tasks}, (ii) \emph{general-purpose code benchmarks} used to assess functional correctness, (iii) \emph{efficiency-oriented benchmarks} that focus on runtime or resource usage, and (iv) \emph{low-level benchmarks} aimed at testing hardware-aware or compiler-level optimizations.

\emph{Algorithmic tasks} are frequently used to evaluate LLM-generated code, as they reflect competitive programming problems that test algorithmic reasoning and implementation skills. However, the number and selection of tasks vary considerably across studies. Some focus on real-world-inspired problems, while others rely on diverse or hand-curated subsets. For instance, Shehab et al.~\cite{shehab_evaluating_2024} evaluate performance on 31 real-world-inspired tasks, Vartziotis et al.~\cite{vartziotis_learn_2024} select six algorithmic problems, and Tuttle et al.~\cite{tuttle_can_2024} use eight curated examples. Cappendijk et al.~\cite{cappendijk_generating_2024} similarly adopt LeetCode-style tasks to assess optimization under energy constraints. Peng et al.~\cite{peng_large_2024} use selected programs from the Energy Languages benchmark suite~\cite{ian_gouy_computer_nodate} to assess the effectiveness of prompting strategies in generating energy-efficient code. 

Standard \emph{general-purpose code benchmarks} such as HumanEval~\cite{chen_evaluating_2021} and MBPP~\cite{austin_program_2021}, along with their enhanced versions, are also reused in several studies~\cite{huang_effi-code_2024,qiu_how_2025,liu_evaluating_2024}. However, they are primarily employed to establish baseline performance, as they do not include optimization-specific objectives or profiling metrics.

Beyond general-purpose code benchmarks, a growing number of studies employ \emph{efficiency-oriented benchmarks} specifically designed to evaluate performance trade-offs such as runtime and memory usage. Examples include EffiBench~\cite{huang_effibench_2024}, EvalPerf~\cite{liu_evaluating_2024}, PIE~\cite{shypula_learning_2023}, and Effi-Code~\cite{huang_effi-code_2024}. Some of these datasets, such as PIE and Effi-Code, serve both training and evaluation purposes. ENAMEL~\cite{qiu_how_2025} also contributes to this space by constructing an energy-aware benchmark from HumanEval and HumanEval+ examples.

At the compiler level, Rosas et al.~\cite{rosas_should_2024} employ \emph{low-level benchmarks} which incorporate established parallel computing suites such as the NAS Parallel Benchmarks~\cite{david_h_bailey_john_t_barton_thomas_a_lasinski_horst_d_simon_nas_1991} and PolyBench~\cite{t_yuki_l_n_pouchet_polybenchc-421_nodate}, alongside their own PCB Suite, to evaluate low-level performance characteristics and test LLMs on optimization patterns traditionally used to assess classical compilers.

To summarize, recent studies employ a variety of benchmarks that differ in task design, evaluation focus, and application context. These range from algorithmic tasks and functional correctness benchmarks to performance-oriented and compiler-level evaluations, reflecting the diverse objectives in assessing LLM-generated code.

\begin{table}[htbp]
\centering
\caption{Benchmarks Used Across Studies}
\label{tab:benchmarks}
\begin{tabular}{@{}>{\raggedright\arraybackslash}p{2cm}
                >{\raggedright\arraybackslash}p{3cm}
                >{\raggedright\arraybackslash}p{2cm}@{}}
\toprule
\textbf{Type} & \textbf{Benchmarks} & \textbf{References} \\
\toprule

Algorithmic Tasks
& \begin{minipage}[t]{\linewidth}\begin{itemize}[leftmargin=*]
  \item Selected LeetCode-style problems (E)
  \item Energy-Language (E)
\end{itemize}\end{minipage} 
& \cite{cursaru_controlled_2024,cappendijk_generating_2024,vartziotis_learn_2024,tuttle_can_2024,shehab_evaluating_2024, peng_large_2024} \\
\midrule
General-purpose Code Benchmarks 

&\begin{minipage}[t]{\linewidth} \begin{itemize}[leftmargin=*]
  \item HumanEval (E)
  \item HumanEval+ (E)
  \item MBPP+ (E)
\end{itemize}\end{minipage}  
& \cite{huang_effi-code_2024,huang_effibench_2024,qiu_how_2025,liu_evaluating_2024} \\
\midrule
Efficiency Benchmarks 
& \begin{minipage}[t]{\linewidth}\begin{itemize}[leftmargin=*]
  \item EffiBench (E)
  \item EvalPerf (E)
  \item ENAMEL (E)
  \item Effi-Code (E,T)
  \item PIE (E,T)
\end{itemize}\end{minipage} 
& \cite{huang_effibench_2024,huang_effi-code_2024,liu_evaluating_2024,qiu_how_2025,duan_leveraging_2023} \\
\midrule
Compiler-Level Benchmarks 
& \begin{minipage}[t]{\linewidth}\begin{itemize}[leftmargin=*]
  \item NAS Parallel (E)
  \item Polybench (E)
  \item PCCB (E)
\end{itemize}\end{minipage} 
& \cite{rosas_should_2024} \\

\bottomrule
\end{tabular}%

\vspace{2pt}
\begin{flushleft}
\footnotesize (T) indicates that the benchmark is used for training purposes; (E) indicates its use for evaluation.
\end{flushleft}
\end{table}

\subsubsection{Evaluation Metrics}
\label{ee-code-optimization-evaluation-metrics}
To evaluate performance-oriented improvements in LLM-generated code, most studies rely on a combination of traditional performance metrics and more specialized efficiency indicators. The most commonly reported metrics include execution time and memory usage. Some works go further by reporting energy consumption either in joules or millijoules. Several studies introduce composite metrics that combine multiple measures. For instance, Vartziotis et al.~\cite{vartziotis_learn_2024} propose Green Capacity, which quantifies improvements across runtime, memory, energy, and \glspl{FLOP} while factoring in correctness. Qiu et al.~\cite{qiu_how_2025} introduce eff@k a runtime based generalization of pass@k that accounts for execution-time censorship and sample diversity. Similarly, Liu et al.~\cite{liu_evaluating_2024} apply the Differential Performance Score (DPS), which profiles optimized code under performance-stressing inputs.

Profiling tools and methodologies vary significantly. Runtime profilers like Perf~\cite{vartziotis_learn_2024,liu_evaluating_2024} and hardware interfaces such as \gls{RAPL}~\cite{peng_large_2024} are commonly used to capture efficiency metrics. Some studies use 
tools like pyJoules~\cite{tuttle_can_2024}, the Monsoon Power Monitor~\cite{cursaru_controlled_2024}, or the CaRV replay framework~\cite{rosas_should_2024}, to measure power usage or consistency of execution. Static analysis tools like Lizard are also used to assess maintainability and complexity~\cite{shehab_evaluating_2024}. In more structured workflows, Liu et al.~\cite{liu_evaluating_2024} introduce Differential Performance Evaluation (DPE) which is a methodology that clusters solutions and employs performance-stressing input generation to rigorously benchmark optimized code across tasks.

Beyond runtime, several studies emphasize different metrics. Peng et al.~\cite{peng_large_2024} consider latency, while Shehab et al.~\cite{shehab_evaluating_2024} analyze code quality using cyclomatic complexity and maintainability index. Cappendijk et al.~\cite{cappendijk_generating_2024} focus on code similarity before and after optimization as a proxy for structural change. In contrast, correctness is typically evaluated with test suites~\cite{tuttle_can_2024}, pass@k scoring~\cite{huang_effibench_2024, liu_evaluating_2024}, or strong test case generators~\cite{qiu_how_2025}.

Overall, the reviewed literature demonstrates metric diversity. However, evaluation pipelines remain heterogeneous in profiling depth, correctness assurance, and evaluation scope.

Table~\ref{tab:metrics} summarizes the evaluation metrics employed across studies, categorizing them based on the type of performance or quality indicator applied. Metric categories in Table~\ref{tab:metrics} are assigned based on how each study utilizes or derives from the corresponding measure. For example, studies listed under \emph{Runtime} either report raw execution time, use normalized variants, or build composite efficiency metrics that rely on runtime data such as Green Capacity~\cite{vartziotis_learn_2024}, DPS~\cite{liu_evaluating_2024}, and eff@k~\cite{qiu_how_2025}, RuntimeRatio~\cite{tuttle_can_2024}. This classification helps reflect both direct use and methodological dependence as a core performance indicator.

\begin{table}[htbp]
\centering
\caption{Categorization and Frequency of Evaluation Metrics Across Studies}
\label{tab:metrics}
\footnotesize
\begin{tabular}{lcl}
\toprule
\textbf{Metric Category} & \textbf{\# of References} & \textbf{References} \\
\toprule

Correctness 
& 12
& \cite{shehab_evaluating_2024,tuttle_can_2024,huang_effi-code_2024,huang_effibench_2024,liu_evaluating_2024,duan_leveraging_2023,qiu_how_2025,rosas_should_2024,peng_large_2024,vartziotis_learn_2024,cursaru_controlled_2024,cappendijk_generating_2024} 
\\
\midrule
Runtime 
& 10
& \cite{shehab_evaluating_2024,tuttle_can_2024,huang_effi-code_2024,huang_effibench_2024,duan_leveraging_2023,rosas_should_2024,vartziotis_learn_2024,cappendijk_generating_2024,liu_evaluating_2024,qiu_how_2025} \\
\midrule
Energy Consumption 
& 5 
& \cite{vartziotis_learn_2024,tuttle_can_2024,peng_large_2024,cappendijk_generating_2024,cursaru_controlled_2024} \\
\midrule
Memory Usage 
& 5
& \cite{shehab_evaluating_2024,huang_effibench_2024, huang_effi-code_2024,cappendijk_generating_2024,vartziotis_learn_2024} \\
\midrule
\glspl{FLOP} 
& 2 
& \cite{vartziotis_learn_2024,cappendijk_generating_2024} \\
\midrule
Code Quality 
& 1 
& \cite{shehab_evaluating_2024} \\
\midrule
Latency 
& 1 
& \cite{peng_large_2024} \\

\bottomrule
\end{tabular}
\end{table}

\subsubsection{Training Setup}
\label{ee-code-optimization-training-setup}
Among the reviewed studies, two studies use fine-tuning to improve code efficiency during training: EffiCode~\cite{huang_effi-code_2024} and PerfRL~\cite{duan_leveraging_2023}. While both aim to enhance functional correctness and runtime performance, they adopt distinct methodologies for data curation and training design.

EffiCode follows a supervised fine-tuning approach using a curated dataset built from open-source datasets. Each code task is paired with automatically generated test cases and profiled using runtime and memory metrics. These profiling signals are integrated into a multi-round optimization framework, SOAP, which refines code based on performance bottlenecks. The final dataset comprises performance optimized solutions, which are used to fine-tune \glspl{LLM}. The training prompt format is instruction-based, without the use of \gls{ICL}. EffiCode is evaluated on HumanEval and EffiBench, providing both correctness and efficiency measurements across a range of code generation tasks.

PerfRL combines supervised learning with \gls{RL} to optimize code using execution-based feedback. It is trained on the PIE dataset, which consists of annotated (slow, fast) code pairs reflecting real-world performance edits. The training input for RL consists of a natural language instruction concatenated with the original (slower) code. During the RL phase, the model generates candidate solutions which are executed and scored based on unit test outcomes and runtime performance. A ranking-based reward mechanism is applied to guide model updates toward more efficient and correct outputs. Both training and evaluation are conducted on PIE dataset.

\section{Discussion}
\label{discussion}

This section synthesizes the key findings from our analysis and focuses on addressing our research questions (Section~\ref{addressing-rqs}), as defined in Table \ref{tab:RQs} and guiding this review. We also reflect on the limitations and threats to validity of this study (Section~\ref{limitations-threats-validity}).

\subsection{Summary of Findings}
\label{addressing-rqs}
\subsubsection{RQ1: Classification and Quantification of Sustainability in ICT Systems}
In analyzing the literature related to \hyperlink{RQ1}{RQ1}, we found that sustainability assessment in \gls{ICT} systems can be organized along two orthogonal axes: component granularity and metric domain. The components considered include hardware (e.g., racks, servers, individual CPUs), software, and network elements, while the metric domains comprise power/energy, efficiency, and environmental impact.

In the hardware domain, power/energy and efficiency metrics dominate, see Tables~\ref{tab:power-energy-classification} and \ref{tab:energy-efficiency-metrics}. The majority of these metrics report instantaneous power draw, total energy consumption, or throughput-based efficiency (e.g., FLOPS per Watt, \gls{IOPS} per Watt~\cite{kipp_layered_2012}). Environmental metrics appear far less frequently and usually reduce to a direct CO$_2$ conversion of the underlying energy data (e.g.,~\cite{priya_survey_2013}). The metric granularity ranges from fine-grained on-chip structures (CPUs, DRAM, GPUs) through to system-of-systems indicators such as \gls{PUE} and \gls{DCIE} that cover data centers or on-premises infrastructures. Most metrics concentrate on the two extremes, the CPU component and system-of-systems, as indicated in Table \ref{tab:energy-efficiency-metrics}.

Similarly, in the software domain, power/energy and efficiency metrics are the primary focus. Granularity ranges from single functions to whole applications and complete systems (e.g., virtual machines or containers). The metrics themselves are not tied to a specific granularity level; they instead compare different software versions, often an optimized version versus a baseline version (e.g. Greenup \cite{green_software_foundation_gps-up_2021}).

At the network level, only one concrete metric was identified, highlighting the need for further research in this area.

Across the reviewed literature, metric measurement focuses primarily on the hardware and software layers (Table \ref{tab:overview-pee-ee-ei}). 

To obtain power and energy values for hardware components, researchers commonly use on-device interfaces provided by the hardware itself, such as Intel \gls{RAPL}~\cite{jiang_edge_2019, guldner_development_2024} and nvidia-smi \cite{guldner_development_2024}. Hardware manufacturers have progressively integrated these interfaces over the years. For example, modern GPUs routinely expose power or energy readings, and some recent CPU platforms provide comparable energy telemetry.

For software, tools in the power/energy domain are the most frequently studied, as shown in Table \ref{tab:overview-pee-ee-ei}, especially for whole software elements. Fewer tools exist for energy efficiency and environmental-impact metrics targeting software elements. Typically, these tools employ model-based estimation, i.e., formulas that transform captured data points into an output metric such as energy (e.g. pTop \cite{noureddine_review_2013}, Kepler ~\cite{apiumhub_18_2024}). Consequently, fine-grained elements, such as function calls or small code snippets, are studied less often because collecting high-resolution data points is complex or sometimes not feasible (e.g. JoularJX ~\cite{apiumhub_18_2024}). We also note that tools (e.g. CodeCarbon ~\cite{CodeCarbon}) addressing energy efficiency and environmental impact depend heavily on power/energy measurements. Hence, their accuracy has a significant impact on the plausibility of the derived metrics.

Finally, no gold standard yet exists for the set of data points for tools to use when estimating power or energy. \cite{guldner_development_2024} have taken a step towards such a standard by proposing a reference model to determine the environmental impact of software.

In summary, sustainability assessment in the \gls{ICT} sector can be organized along the two axes of granularity and metric domain. Most literature addresses the hardware and software layers within the power/energy domain. Conversely, energy efficiency and environmental-impact metrics remain relatively under-explored at all granularities.

\subsubsection{RQ2: Optimization of Energy Efficiency and Power Consumption in ICT Systems, specifically Data Centers and Cloud Computing Environments}
Recent literature across both academia and industry identifies a wide range of software- and system-level approaches, methods, tools and techniques in relation to \hyperlink{RQ2}{RQ2}. As previously shown, a central and long-standing approach involves dynamic power and resource management techniques such as \gls{DVFS}, adaptive task scheduling, and workload balancing. These methods allow systems to adjust resource usage in real time based on factors such as workload, energy availability, and sustainability constraints. They are often  aligned with \glspl{SLA} to maintain performance guarantees~\cite{khattar_toward_2019,priya_survey_2013,katal_energy_2023}. Within this category, a new area of interest is the tighter integration with renewable energy sources, such as adjustments driven by green energy availability \cite{li_iswitch_2012}.

Another key area is energy-efficient \gls{VM} and container management. These server consolidation techniques aim to minimize the number of active physical machines by intelligently migrating and placing \glspl{VM} or containers, thereby reducing idle energy consumption~\cite{hosseini_shirvani_survey_2020}. These optimizations are typically realized using operational metrics such as workload and component utilization. More advanced techniques incorporate workload correlation analysis and heuristic optimization to balance  energy savings with migration overhead, uptime, and \gls{SLA} adherence~\cite{ahmad_virtual_2015,azizi_energy-efficient_2020,yakhchi_proposing_2015,zheng_joint_2014}. In recent years, container-based approaches have garnered increased attention due to their flexibility and reduced overhead compared to traditional \gls{VM} solutions~\cite{gholipour_novel_2020}. Collectively, the reviewed works emphasize that achieving substantial reductions in energy consumption in \glspl{DCCCE} increasingly depends on multifaceted approaches that combine monitoring, prediction, optimization, and hybrid virtualization techniques~\cite{yang_implementation_2020,choudhary_critical_2016,hsu_optimizing_2014}.

Broadening the perspective, the systematic incorporation of energy-awareness into software engineering practices and system architecture design has emerged as an important research direction. Methods such as process-level power profiling, workload-specific resource scheduling, and energy-aware coding practices can significantly influence energy consumption. Empirical studies indicate that software developers' decisions, ranging from algorithm selection to memory management, can exert a direct influence on the energy consumption and efficiency of \gls{ICT} systems. Consequently, empowering developers to embed sustainability considerations into the software development life cycle is essential.

From a system-wide and macro-level perspectives, recent literature increasingly advocates cross-layer coordination to improve energy efficiency across the full stack of physical and virtual infrastructure. Research in this area frequently leverages layered energy models, green metrics, and staged optimization frameworks. These methods support the operationalization of energy- and sustainability-awareness and enable concrete, measurable interventions.

Despite the diversity of approaches, several limitations remain. One key limitation is the lack of integrated, cross-layer optimization strategies that harmonize energy management across software, system, and infrastructure levels~\cite{kanso_review_2024,uchechukwu_energy_2014}. Moreover, empirical validation is often conducted in simulated environments, and many studies provide limited evidence from real-world deployments. Future research should prioritize holistic architectures that unify energy optimization across all layers of \gls{DCCCE}~\cite{sofia_implementation_2015,manimegalai_energy_2024}. Advancements in machine learning and \glspl{LLM} also hold promise for predictive energy consumption modeling and dynamic balancing of workloads, \glspl{VM}, and containers. Finally, there is a need to embed energy-awareness and related metrics into software development environments and system design processes~\cite{johnston_improving_2023,meinhardt_software_2023}.

\subsubsection{RQ3: Applications of LLMs for Automating Code Generation and Transformation Toward Resource-Efficient Software Systems}
Where \hyperlink{RQ2}{RQ2} addressed optimization at the system, orchestration, and design level, \hyperlink{RQ3}{RQ3} examines a complementary and emerging layer: the code itself. Recent literature investigates a variety of \gls{LLM}-based approaches to automate code generation and transformation tasks with the goal of improving resource efficiency and software sustainability, as examined in Section \ref{ee-code-optimization}. As shown in Table~\ref{tab:contribution-task}, most reviewed works adopt a code generation setting, where models are prompted to synthesize implementations from natural language descriptions (e.g.~\cite{liu_evaluating_2024,cappendijk_generating_2024,tuttle_can_2024}). In contrast, only a few studies focus on code transformation, where models are used to optimize existing code for performance improvements (e.g. ~\cite{duan_leveraging_2023,peng_large_2024,rosas_should_2024}). This imbalance suggests that the potential of \glspl{LLM} to support software maintenance and refactoring remains underexplored.

Prompting strategies vary widely across studies, as outlined in Table~\ref{tab:prompts}. While some works employ basic or objective-driven prompts (e.g. ~\cite{tuttle_can_2024,cursaru_controlled_2024}), others explore iterative prompting and feedback-based refinement (e.g.~\cite{peng_large_2024}). However, most studies primarily evaluate prompt effectiveness empirically, with comparatively limited attention to developing end-to-end, integrated LLM-based optimization methods.

Another recurring pattern is the dominance of Python as the target language. As seen in Table~\ref{tab:pls}, Python appears in most studies, while performance-critical languages such as C, C++, or Java are underrepresented (e.g.~\cite{shehab_evaluating_2024,rosas_should_2024}). This prevalence likely reflects the Python-centric nature of current LLM research ecosystems and the frequent use of Python as a convenient research proxy in experimental evaluation. However, this limits the generalization of current findings, especially for domains such as embedded systems or high-performance computing.

Evaluation methodologies are highly heterogeneous across studies both in terms of the metrics used and the depth of profiling instrumentation. While execution time and memory usage are frequently reported (e.g.,~\cite{cappendijk_generating_2024,shehab_evaluating_2024}), only a limited number of works directly measure energy consumption (e.g.,~\cite{cursaru_controlled_2024,cappendijk_generating_2024}). As discussed in Section~\ref{ee-code-optimization-evaluation-metrics}, studies employ a wide range of performance-related indicators, including runtime, memory usage, and energy consumption. Several studies introduce derived or composite metrics such as Green Capacity~\cite{vartziotis_learn_2024}, DPS~\cite{liu_evaluating_2024}, and \textit{eff@k}~\cite{qiu_how_2025}. However, even when similar metrics are adopted, studies often differ in how they represent and interpret them. For example, memory usage may be reported as raw, peak, total, or normalized values, while runtime may be expressed in absolute terms or as relative ratios such as speedup, runtime reduction, or normalized execution time. In addition, measurement intervals and profiling tools differ across studies. This inconsistency complicates direct comparison and limits reproducibility. These challenges reflect broader concerns raised under Section \ref{ee-measurement}, where the absence of unified evaluation frameworks and benchmarking protocols similarly hinders the ability to quantify or compare optimization outcomes. 

In summary, while interest in using \glspl{LLM} for resource-efficient code generation and transformation is increasing, the field still faces several key limitations. These include an under-representation of code transformation tasks, limited methodological innovation beyond prompting, a strong bias towards Python, and inconsistent evaluation pipelines. Addressing these gaps will be essential to improve reproducibility, comparability, and the practical relevance of future work in resource-aware \gls{LLM}-assisted software development.

\subsubsection{Toward a Minimal Reporting Checklist}
\label{minimal-reporting-checklist}
Across the reviewed literature, comparability is limited more by inconsistent reporting than by a lack of available metrics. We recommend that software-energy studies in this space report at minimum: (1) energy or power in explicit units at a stated granularity (cf. Table~\ref{tab:power-energy-classification}); (2) an explicit baseline, since efficiency claims are only meaningful relative to a stated reference; (3) a decomposed efficiency metric (e.g., Speedup/Powerup/Greenup~\cite{green_software_foundation_gps-up_2021} rather than \gls{EDP} alone), to make explicit whether gains stem from runtime, power, or both; (4) for environmental-impact claims, the grid carbon-intensity source and system boundary used; and (5) for LLM-assisted code tasks, correctness reported alongside efficiency. No reviewed source reports all five consistently, underscoring the comparability gap across \hyperlink{RQ1}{RQ1}-\hyperlink{RQ3}{RQ3}.

\subsection{Limitations and Threats to Validity}
\label{limitations-threats-validity}
To ensure repeatability and transparency, we employed a  systematic approach as detailed in Section \ref{methodology}.  Throughout the research process, we identified potential limitations and threats to validity and took steps to address and mitigate them. Several points are noteworthy and described in the following section.

The chosen research scope is deliberately broad, encompassing various aspects of energy measurement, optimization and the influence of code generation and transformation of energy efficiency. While this wide-ranging approach enables a holistic overview of the relevant domains, it may come at the cost of reduced depth in individual subdomains. This trade-off positions the review as a foundational synthesis that can inform and guide the development of more focused or complementary work in the field and support the identification of synergies across subdomains.

Relying on Google Search and Google Scholar, each capped at the first 50 results per query, bounds the search to a practical, reproducible scope but does not guarantee representativeness: search-engine ranking is influenced by factors unrelated to topical relevance (e.g., popularity, search engine optimization), and a fixed per-query cutoff may exclude relevant sources ranked lower. Complementing this search with dedicated academic databases (e.g., Scopus, IEEE Xplore, ACM Digital Library) could surface additional peer-reviewed sources missed by the current strategy; we leave this to future replications.

Regarding the search and analysis process, primary sources were analyzed systematically, beginning with the definition of thematic codes and classification variables, which were subsequently populated with extracted data. Although this structured approach enhances consistency, it may omit information that does not fit into the predefined codes and variables.

Further, the analysis was conducted without redundant parallel coding or cross-validation among multiple reviewers. As a result, some interpretations may reflect subjective perspectives of the individual researcher, potentially affecting the reliability of the results.

Our findings are structured along different conceptual categories. While measurement-related topics are categorized according to targeted elements (software, hardware, and networks), optimization strategies are grouped into broader conceptual clusters (General optimization in \gls{ICT} systems, in particular in \glspl{DCCCE}, and \glspl{LLM}-based approaches for code generation and transformation). This structural asymmetry, while intentional for clarity, may limit direct comparability between measurement and optimization domains.

Given the multi-vocal nature of the review and the inclusion of non-peer-reviewed literature, variability in rigor across the included sources must be recognized and carefully addressed. 
Gray literature and all non-peer-reviewed formal literature sources (such as bachelor's and master's theses) were evaluated using the quality assessment framework described in Section~\ref{methodology} to ensure relevance and reliability before inclusion.

While gray literature was included to enhance comprehensiveness, many of the retrieved sources,  primarily due to the broadness of our search terms, were high-level in nature. They often provided overviews or general recommendations without adding actionable or context-specific insights beyond what was covered in the formal literature. Consequently, their analytical value was limited in practice.

\section{Conclusions and Future Work}
\label{conclusion}
The opening section of our results analyses how sustainability is currently assessed in ICT systems. It first identifies and classifies core sustainability-related metrics and then surveys the measurement approaches and tools used to capture these metrics across the hardware, software, and network layers. Our review shows that the vast majority of published work concentrates on power/energy and efficiency metrics for hardware and software elements, and that tooling support is strongest in this space. However, acquiring hardware power data still depends on proprietary APIs. Thus, a vendor-agnostic interface has yet to emerge. Moreover, no gold standard defines the data points required to derive power/energy, energy efficiency, or broader sustainability metrics for software. The network layer is even less studied, with only a few metrics and tools available. Future research should therefore prioritize vendor-neutral telemetry for hardware, a validated, community-accepted data model for software-level metrics, and an expanded metric and measurement toolkit for network components.

The second section of our results focused on optimization in \gls{ICT} systems, with a specific focus on \glspl{DCCCE}, and explored a broad landscape of optimization strategies, methods, metrics, and tools applied at varying levels of granularity. To provide structure and clarity, we synthesized these diverse sources into coherent and logically connected thematic groupings, highlighting key layers of interest and potential within the literature. However, current research remains largely fragmented. Many contributions address narrowly defined problems (e.g., specific VM migration strategies or resource scheduling algorithms) and often lack integration across the full system stack. Future work should therefore prioritize unified, cross-layer energy optimization frameworks. It should also enable developers to address sustainability concerns proactively by integrating energy-awareness into software engineering practices, tools, and decision-making processes.

The final section of our results reviewed how \glspl{LLM} are applied to automate code-level generation and transformation with the aim of improving the resource efficiency of software systems. To make sense of this rapidly evolving and multifaceted research direction, we synthesized findings from recent studies across multiple dimensions, including task formulation, prompting strategies, programming language focus, benchmarks, evaluation metrics, and training setups. The reviewed studies indicate that most works emphasize code generation from natural language using foundation models, while code transformation tasks and approaches involving model fine-tuning or structured workflows are less frequently explored. A variety of prompting strategies and LLM pipelines are investigated across studies, reflecting diverse design choices in how tasks are formulated and executed. However, evaluation methodologies differ considerably in terms of metrics, tools, and profiling depth. This complicates cross-study comparison and limiting reproducibility. Notably, Python remains the dominant language of focus, with relatively little work addressing performance-critical contexts such as C, C++, or Java. Expanding coverage to include transformation-based tasks, improving consistency in evaluation practices, and exploring novel modeling frameworks are important directions for future work in this area.

In summary, our findings reveal a growing interest in detailed energy and sustainability measurement, cross-layer energy optimization, energy-aware software engineering, and the use of \glspl{LLM} to improve code-level sustainability, while also highlighting notable research gaps. Across all areas, we identified both emerging opportunities and persistent fragmentation, particularly in terms of inconsistent metrics, limited integration across system layers, and limited reproducibility. These insights underscore the need for future research to develop holistic, interoperable frameworks that combine system-level energy optimization with developer-centric tools and methods. Additionally, there is significant potential in advancing LLM-based techniques beyond code generation, especially in transformation-based tasks, support for diverse programming languages, and robust, reproducible evaluation protocols.

\section*{Acknowledgment}
This work was supported by ITEA4 and the Eureka Cluster on software innovation project "23016 GreenCode" supported with funding from the Federal Ministry of Research, Technology and Space (BMFTR), Germany (grant number 16IS24070G).
The authors used GPT-4o (OpenAI) and Claude Sonnet 4.6 (Anthropic) during manuscript preparation for language editing and wording suggestions.

\bibliographystyle{IEEEtran}
\bibliography{references}

\end{document}